\documentclass[prd,twocolumn,showpacs,showkeys,preprintnumbers,floatfix,
nofootinbib,superscriptaddress]{revtex4-1}
\usepackage{amsfonts} 
\usepackage{amssymb} 
\usepackage{amsmath} 
\usepackage{graphicx} 
\usepackage{subfigure} 
\usepackage{array} 
\usepackage{dcolumn} 
\usepackage{bm} 
\usepackage{latexsym} 
\usepackage{longtable} 
\usepackage[colorlinks=true,linkcolor=blue,citecolor=blue,filecolor=blue,urlcolor=black]{hyperref} 
\usepackage[all]{hypcap} 

\graphicspath{{./figs/}}
\renewcommand{\tilde}{\widetilde}

\newcommand{\CO}{{\cal O}}

\newcommand{\CU}{{\cal U}}

\newcommand{\CG}{{\cal G}}

\renewcommand{\bar}{\overline}
\renewcommand{\hat}{\widehat}
\renewcommand{\tilde}{\widetilde}

\newcommand{\gambar}[1]{ \overline{\overline{(\gamma_{#1}\otimes I)}} }

\newcommand{\ixibar}[1]{ \overline{\overline{(I\otimes\xi_{#1})}} }

\newcommand{\singlebar}[2]{ {\overline{(\gamma_{#1}\otimes\xi_{#2})}} }
\newcommand{\doublebar}[2]{ \overline{\overline{(\gamma_{#1}\otimes\xi_{#2})}} }
\newcommand{\semitimes}{\mathrel>\joinrel\mathrel\triangleleft}

\newcommand{\tr}{\textrm{tr}}
\newcommand{\Tr}{\textrm{Tr}}

\begin{document}


\title{Non-perturbative Renormalization for
Improved Staggered Bilinears}

\author{Andrew T. Lytle}
\email[Email: ]{atlytle@theory.tifr.res.in}
\affiliation{
Department of Theoretical Physics, Tata Institute of Fundamental Research, 
 Mumbai 400005, India \\
}
\author{Stephen R. Sharpe}
\email[Email: ]{srsharpe@uw.edu}
%
%
\affiliation{
 Physics Department, University of Washington, 
 Seattle, WA 98195-1560, USA \\
}

\date{\today}
\begin{abstract}

We apply non-perturbative renormalization to bilinears composed of 
improved staggered fermions.
We explain how to generalize the method to staggered fermions in
a way which is consistent with the lattice symmetries, and introduce
a new type of lattice bilinear which transforms covariantly and
avoids mixing. We derive the consequences of lattice symmetries for
the propagator and vertices.
We implement the method numerically for hypercubic-smeared (HYP) and
asqtad valence fermion actions, 
using lattices with asqtad sea quarks generated by the MILC collaboration.
We compare the non-perturbative results so obtained
to those from perturbation theory, using both scale-independent ratios of
bilinears (of which we calculate 26), 
and the scale-dependent bilinears themselves.
Overall, we find that one-loop perturbation theory provides a successful
description of the results for HYP-fermions if we allow for a
truncation error of roughly the size of the square of the one-loop term
(for ratios) or of size ${\cal O}(1)\times\alpha^2$ (for the bilinears
themselves).
Perturbation theory is, however, less successful at describing the
non-perturbative asqtad results.
\end{abstract}
%
%
\maketitle

\section{Introduction \label{sec:intr}}
Precise knowledge of matching factors ($Z$-factors) 
between lattice operators and
their continuum counterparts is necessary for many phenomenological
applications of lattice QCD. 
Non-perturbative renormalization (NPR)~\cite{NPR}
is a widely used method for determining these matching factors,
and has been applied successfully to many types of lattice fermion.\footnote{%
For a review see Ref.~\cite{yaokireview}.}  
Compared to perturbative matching, which is necessarily carried out
at fixed order, NPR has the great advantage of avoiding truncation errors. 
While the size of such errors can be estimated, the estimates are
necessarily approximate. 

In this article we apply NPR to improved staggered fermions, focusing
on matching factors for quark bilinears. There have been relatively
few applications of NPR to staggered fermions. Most relevant for
our work is a quenched calculation of $Z_m$, 
the renormalization factor for the quark mass,
using unimproved staggered fermions~\cite{ishi}.
This calculation found large discretization errors, which is typical
for unimproved staggered fermions. Such errors should be significantly
reduced by using improved actions as we do here.

Generalizing NPR to staggered fermions is relatively straightforward,
although there are a number of technical details that do not arise with
Wilson-like fermions and have not been discussed in previous work. 
We apply the method to the quark propagator and to quark bilinears
having arbitrary spin and taste (but no derivatives).
We use two types of improved staggered quarks: 
``asqtad''~\cite{asqtada,asqtadb,asqtadlepage} 
and HYP-smeared improved staggered quarks~\cite{HYP}.
A complication arising with staggered quarks is the 
presence of the taste degree of freedom, 
which has the consequence that each bilinear comes in 16 possible tastes.
In this study we turn the presence of multiple tastes into an advantage.
Ratios of matching
factors having the same spin but different tastes become unity in the
continuum limit, but differ at finite lattice spacing.\footnote{%
That all ratios (including those containing the taste singlet)
become unity in the continuum limit
holds only because we consider flavor non-singlet bilinears, 
for which there are no quark-disconnected Wick contractions.}
The differences are proportional to both ${\cal O}(\alpha_s)$ 
(with the coupling evaluated at a scale $\sim 1/a$) 
and $a^2 p^2$ (where $p$ is the scale at which NPR is implemented). 
Such ratios are akin to $Z_A/Z_V$ with Wilson fermions.
Comparing them to the results from one-loop perturbation theory (PT), 
and studying their $p$ dependence gives information on the
accuracy of truncated PT and may allow discretization effects 
and perturbative contributions to be disentangled.
The multiple tastes of staggered fermions allow us to form
many such ratios.
We also make the comparison with PT for the $Z$-factors themselves.\footnote{%
We do not present results
for the (spin) pseudoscalar in this article,
since for these quantities
the chiral limit is complicated by the presence of pion poles.  
This is discussed further in Sec.~\ref{subsec:vertex}.}

Our work was initially
motivated by the need for matching factors in two ongoing
calculations---that of quark masses by the MILC collaboration
and of electroweak matrix elements by the SWME collaboration.
The former work has
determined the light quark masses using first 
the asqtad action~\cite{asqtada,asqtadb,asqtadlepage}
(with results reviewed in Ref.~\cite{milcrmp})
and more recently the HISQ action~\cite{hisq} 
(with results exemplified by those of Ref.~\cite{milchisq}).
These determinations use two-loop matching factors, and the concomitant
truncation error is the largest source of error.
A non-perturbative determination of the matching factor could firm up
and reduce this error. It would also allow check of the consistency
of different lattice approaches by comparing with the more precise
results for light quark masses obtained in Ref.~\cite{HPQCDms}
using a combination of results for $m_s/m_c$ and $m_c$.

First results using NPR for the matching factor $Z_m$ with asqtad quarks
were presented by one of us in Ref.~\cite{Lytlelat09} and extended
in Ref.~\cite{Lytlethesis}. Using the MILC coarse ($a\approx 0.12\;$fm)
and fine ($a\approx 0.09\;$fm) lattices to take the continuum limit, 
the result obtained for the strange quark mass was 
$m_s(\overline{\rm MS}, 2\;GeV) =103 \pm 3\;$MeV, where the error is
only statistical. This is somewhat higher than the results
one obtains from these two lattices
using one-loop~\cite{MILC1loop} ($m_s=76\pm 8\;$MeV) 
and two-loop~\cite{MILC2loop} matching ($m_s=87\pm 6\;$MeV).\footnote{%
The result based on four lattice spacings and two-loop matching is 
$m_s=88\pm 5\;$MeV~\cite{milcrmp}.
We also note that the most precise determination, 
(obtained using the ratio $m_s/m_c$)
is $m_s=92.4\pm 1.5\;$MeV~\cite{HPQCDms}.} 
What is needed, however, is a full error budget for the NPR calculation.
One aim of the present work is to study some of the systematic errors
that enter into this budget.

The second ongoing calculation which motivates the present
work is that of $B_K$ (and related matrix elements) using HYP-smeared
staggered fermions on the MILC asqtad configurations~\cite{SW-1,SW11}. 
This calculation uses one-loop matching for the relevant
four-fermion operators,
and the truncation error again dominates that from other sources.
This error can be significantly reduced using NPR.
The present calculation is a step on the way,
as the four-fermion operators are 
essentially composed of products of the bilinears studied here.

For completeness, we recall the main disadvantages of NPR.
These are the
need for a ``window'' where non-perturbative and discretization 
errors are small,
the presence of statistical errors,
and the possibility of ``Gribov noise''. 
Methods exist, however, to systematically reduce the first two errors.
The window can be enlarged by combining the step-scaling technique
with NPR~\cite{Arthur:2010ht} (a technique we do not use here), and 
statistical errors can be substantially reduced 
using momentum sources (which we do use).
Gribov noise\footnote{%
Gribov noise can be avoided using methods based on the Schr\"odinger
functional~\cite{SchrFa,SchrFb},
but these are more complicated in practice than NPR.} 
is the uncertainty caused by the
presence of multiple solutions to the gauge-fixing 
criterion~\cite{gfix1a,gfix1b,gfix1c,gfix2,gfix3a,gfix3b}.

This paper is organized as follows.
The following section describes the application of
NPR to staggered fermions, beginning with the quark propagator
and then discussing bilinears.
We introduce and use ``covariant bilinears'',
 which transform in irreducible representations
(irreps) of the lattice symmetry group, 
and differ somewhat from the
``hypercube bilinears'' commonly used in simulations.
In Sec.~\ref{sec:numerics} we briefly describe the numerical
methods we use and their implementation.
We present our results 
in Sec.~\ref{sec:results}, providing a detailed comparison
with perturbation theory.
We conclude in Sec.~\ref{sec:conc}.

Technical results are collected in four appendices.
Appendix~\ref{app:irreps} sketches the classification of
covariant bilinears into irreps of the lattice symmetry group.
In App.~\ref{app:symm}
we explain how lattice symmetries constrain the form of
the quark propagator and bilinear amplitudes.
In App.~\ref{app:pert} we describe how the perturbative calculation
of one-loop matching factors changes when moving from hypercube
to covariant bilinears.
In App.~\ref{app:running} we collect continuum results needed for the
renormalization scale evolution of the matching factors.

Preliminary results from this study were presented in Refs.~\cite{sharpelat11}
and \cite{andrewlat12}.

\section{NPR for staggered fermions}
\label{sec:nprstagg}

For valence staggered fermions we use either
the unimproved action, the HYP-smeared improved action or
the asqtad action. The unimproved action is
\begin{eqnarray}
S_\text{un} &=&
\sum_{n} \bar{\chi}(n) \Big[
\sum_{\mu} \eta_{\mu}(n) \nabla_\mu + m \Big]\chi(n) \,,
\label{eq:Sunimp}
\\
\nabla_\mu \chi(n) &=&
\frac12
[U_\mu (n) \chi(n + \hat{\mu}) -
U^{\dagger}_{\mu}(n-\hat{\mu}) \chi(n-\hat{\mu})]
\nonumber
\end{eqnarray}
where $\chi(n)$ is the usual single-component staggered lattice field, 
$n = (n_1,n_2,n_3,n_4)$ labels lattice sites,
$\eta_{\mu}(n) = (-1)^{n_1 + \cdots + n_{\mu-1}} $ is the
remnant of the Dirac matrices, 
and $U_\mu(n)$ are the SU(3) gauge links. 
All quantities are dimensionless,
so that, for example, the bare quark mass is related to
the physical mass by $Z_m m = m_{\rm phys} a$.

The HYP-smeared action is obtained simply by replacing the
links with HYP-smeared links, $V_\mu(n)$, obtained as explained
in Ref.~\cite{HYP}. We use the
HYP-smearing parameters labeled ``HYP(1)'' in Ref.~\cite{LSff}:
$\alpha_1=0.75$, $\alpha_2=0.6$ and $\alpha_3=0.3$.

The asqtad action~\cite{asqtada,asqtadb,asqtadlepage} 
is described in App.~\ref{app:notation}.
This action is fully tree-level $O(a^2)$ improved, unlike the
HYP-smeared action where only taste-breaking terms are improved.

We use configurations from the MILC collaboration, which 
are generated with the asqtad action for sea quarks 
(using the rooting prescription to remove unwanted tastes)
and the one-loop improved Symanzik action for gluons~\cite{milcrmp}.
All lattices have an even number of points, $L_\mu$,
in each direction, and we use periodic boundary conditions 
on the propagators in all directions. 

Before calculating propagators and vertices,
gauge fields are fixed to Landau gauge.
On the lattice, this is achieved by maximizing
\begin{equation}
F_L = \sum_{n,\mu} \Tr(U_\mu(n) + U_\mu(n)^\dagger),
\end{equation}
for which we use an overrelaxation algorithm. 
This finds a local maximum, of which there are many, leading to 
the ambiguity of Gribov copies. We simply assume, following standard
practice~\cite{gfix1a,gfix1b,gfix1c,gfix2,gfix3a,gfix3b},
that the differences in the results on different copies
are small enough to ignore.

\subsection{Quark Propagator}
\label{subsec:prop}

NPR takes place in momentum space, so we must choose the
appropriate momentum-space quark fields. 
The choice is non-trivial for staggered fermions,
because the lattice Brillouin zone contains both momentum and
taste information~\cite{GS}. This is the momentum-space analog
of the fact that the four-taste Dirac field is built up from
staggered fields $\chi$ living on a $2^4$ hypercube~\cite{hypercube}.
Motivated by this split into hypercubes,
Ref.~\cite{ishi} used the momentum-space field
\begin{equation}
\phi'_A(p') = \sum_y e^{-ip' \cdot y}\, \chi(y+A),
\label{eq:ishifield}
\end{equation}
where $y$ is a vector labeling $2^4$ hypercubes
($y_\mu = 2 n^y_\mu$, with $n^y_\mu$ integers),
and $A$ is a hypercube vector 
labeling points within the hypercubes ($A_\mu\in \{0,1\}$).
Thus $y+A$ picks out a particular lattice point.
The physical momentum\footnote{%
This momentum is physical in the sense that the part
corresponding to taste degrees of freedom has been removed.
It is, however, dimensionless, containing an implicit factor
of $a$.
}
$p'$ lies in a reduced Brillouin zone,
\begin{equation}
-\pi/2 \le p'_\mu < \pi/2,
\label{eq:subzone}
\end{equation}
and the label $A$ contains the Dirac and taste indices.
The key feature of the choice (\ref{eq:ishifield}) is
that the momentum phase factor does not vary within each
hypercube.

This choice is, however, problematic, because $\phi'_A(p')$ does
not transform irreducibly under lattice translations. 
This is clear from the fact that the division of the
lattice into $2^4$ hypercubes is not invariant under
single-site translations.
The lack of irreducibility implies that the propagator does not
have a simple, continuum-like form.

It is straightforward to avoid this problem by using
the definition introduced by Ref.~\cite{GS}.
One uses the standard Fourier transform, without reference to
hypercubes, leading to a momentum lying in the usual Brillouin zone,
$-\pi \le p_\mu < \pi$. One then breaks this up into $2^4$ subzones,
each characterized by a hypercube vector $B$, such that a general
momentum is written
\begin{equation}
p_\mu = p'_\mu + \pi B_\mu
\,,
\end{equation}
with $p'_\mu$ constrained as in (\ref{eq:subzone}) above.
$p'$ is the physical momentum and $B$ contains the spin and taste
information.
The momentum-space field of Ref.~\cite{GS} is then
\begin{eqnarray}
\phi_B(p') &=& \sum_n e^{-i p \cdot n}\, \chi(n) 
\label{eq:goodmomfield}\\
&=& \sum_{y,A} e^{-i p'\cdot y -ip' \cdot A}\, (-)^{B \cdot A} \chi(y+A)\\
&=& \sum_{A} e^{-ip' \cdot A}\, (-)^{B \cdot A} \phi'_A(p')
\,,
\label{eq:relatemomfields}
\end{eqnarray}
(with an identical definition for $\bar\phi$ in terms of $\bar\chi$).
The second line shows that this new choice differs from
$\phi'_B$ of Eq.~(\ref{eq:ishifield}) by the presence of a
phase factor $\exp(-ip' \cdot A)$ {\em within} the hypercube.
The last line gives the explicit relation between
$\phi_B$ and $\phi'_B$.
In the continuum limit, when one can set $p'\to 0$,
the two fields are simply related by a unitary transformation,
and are thus physically equivalent.
Away from the continuum, however, they differ in an essential way.

The merits of the choice (\ref{eq:goodmomfield}) can be seen by
considering the free quark propagator. 
First we define the propagator (with or without interactions) by
\begin{equation}
\langle \phi_A(p') \bar\phi_B(-q')\rangle
= (2\pi)^4 \overline\delta(p'-q') S(p')_{AB}
\,,
\label{eq:Sdef}
\end{equation}
where $\bar\delta$ is the periodic delta-function
(with period $2\pi$).\footnote{%
On a finite lattice one replaces
$(2\pi)^4\overline\delta(0)$ with 
the number of sites, $N_{\rm site}$.}
This form follows from the 
invariance of the action under two-site translations
without the need for phases on the quark fields.
The propagator $S(p')$ has implicit color indices
and explicit spin-taste indices. Altogether it is
a $48\times 48$ matrix. Invariance under global
gauge transformations implies, however, that it is proportional to
the identity matrix in color space, a property that holds
also for its inverse. Thus we keep color indices implicit
in the following discussion.
For free quarks, the inverse of $S$ is~\cite{GS}
\begin{equation}
S_{\text{free}}^{-1}(p')_{AB} = m\overline{\overline{(1\otimes I)}}_{AB}
+ i \sum_\mu \sin(p'_\mu) \overline{\overline{(\gamma_\mu\otimes I)}}_{AB}
\,,
\label{eq:freeprop}
\end{equation}
where $m$ is the valence quark mass.
Here we use the notation of Refs.~\cite{DK,DS} 
(also briefly explained in App.~\ref{app:notation}).

The result (\ref{eq:freeprop})
has a continuum-like form with a taste-singlet mass term and
a taste-singlet derivative term; the only effect of discretization
is the replacement of $p'_\mu$ with $\sin(p'_\mu)$.
In particular, there are no taste-violating terms.
This simplicity is guaranteed by the lattice symmetries~\cite{GS},
and does not hold if one uses the field (\ref{eq:ishifield}).

In fact, one can show that the absence of taste-violating terms
holds in the presence of interactions.
This was shown in Ref.~\cite{GS} close to the continuum limit, 
and is demonstrated for arbitrary $p'$ in App.~\ref{app:symm}.
The key result is that the propagator satisfies, for each $\mu$,
\begin{equation}
S(p) = \ixibar{\mu} S(p) \ixibar{\mu}
\ \ \Leftrightarrow\ \
[\ixibar{\mu},S(p)]=0
\,.
\label{eq:translateS}
\end{equation}
This implies that
$S(p)$ is a taste singlet, i.e.\ consists only of
terms whose matrix structure is $\gambar{S}$.
We stress that this result holds to all orders in
perturbation theory, and, indeed, non-perturbatively.

Constraints on the form of the propagator also
arise from lattice rotations and spatial inversions,
as discussed in App.~\ref{app:symm}.
Given that only taste-singlet terms appear, however, these
constraints are identical to those that apply to other
types of fermions, e.g.\ Wilson or overlap fermions. 
The final constraints arise from the $U(1)_\epsilon$ axial
symmetry of the staggered action.
The net effect is that the form
of the inverse propagator is\footnote{%
The same constraint applies to the propagator, but for the
NPR procedure it is more convenient to focus on $S^{-1}$.}
\begin{align}
\begin{split}
S^{-1}(p') &= c_S m \overline{\overline{(I\otimes I)}}
+ c_V p'_\mu \overline{\overline{\gamma_\mu\otimes I}}
\\
&{} + c_T m \sum_{\mu\nu} p'_\mu (p'_\nu)^3 \,
\overline{\overline{\gamma_{\mu\nu}\otimes I}}
\\
&{} + c_A \sum_{\mu\nu\rho} p'_\mu (p'_\nu)^3 (p'_\rho)^5\,
\overline{\overline{\gamma_{\mu\nu\rho}\otimes I}}
\\
&{} + c_P m \sum_{\mu\nu\rho\sigma} p'_\mu (p'_\nu)^3 
(p'_\rho)^5 (p'_\sigma)^7\,
\overline{\overline{\gamma_{\mu\nu\rho\sigma}\otimes I}}
\,,
\end{split}
\label{eq:Sinvform}
\end{align}
where the $c_j$ are constants.
Here we are using a somewhat schematic notation in which,
for each Dirac structure,
we display only the term having the lowest power of $p'$ and $m$.
Thus, for example, in the $c_T$ term, there are terms not shown
in which the momentum dependence is $p'_\mu (p'_\nu)^5$, {\em etc.}.
Such terms are suppressed in the continuum limit relative to those shown
by powers of $a^2$.
We are also using the shorthand $\gamma_{\mu\nu}=\gamma_\mu\gamma_\nu$, etc.
The factors of $m$ arise due to the $U(1)_\epsilon$
symmetry (and are thus absent in the corresponding result for
Wilson fermions).
As one approaches the continuum limit (i.e.\ as $p', m \to O(a)$) only the
$c_S$ and $c_V$ terms survive, and one is thus guaranteed to
obtain the same form as the free propagator, up to mass and
wavefunction renormalization.

With this background we can now return to the application
of NPR to staggered fermions. Since the staggered propagator
has the same general form as with other fermions, supplemented
only by the taste degrees of freedom, one can
carry over the formalism of Ref.~\cite{NPR} essentially verbatim.
We first calculate $S(p')$ from Eq.~(\ref{eq:Sdef}),
and then, for each $p'$, invert the resulting $48\times48$ matrix
to obtain $S^{-1}(p')$. 
In the ${\rm RI}'$ scheme, wave-function
renormalization is then given by
\begin{equation}
Z'_q(p') = -i\frac1{48} \sum_\mu \frac{\tilde p'_\mu}{\tilde p^{'2}}
\Tr\left[\gambar{\mu}\; S^{-1}(p')\right]\,.
\label{eq:Zqdef}
\end{equation}
Here $\tilde p'=\sin(p')$ and 
$\sin(p') + \sin^3(p')/6$, respectively,
for HYP and asqtad fermions. These choices are made
so that, for both cases, $Z'_q=1$ in the free theory.
The shorthand $\tilde p^{'2}$ means $\sum_\mu (\tilde p'_\mu)^2$, 
and the trace is over spin, taste and color indices.
As always, with NPR, one aims to work in the window
\begin{equation}
\Lambda_{\rm QCD}^2  \ll p^{'2} \ll \left(\frac{\pi}{a}\right)^2
\,,
\label{eq:npr_condition}
\end{equation}
so as to avoid non-perturbative effects and discretization errors.
We discuss these constraints further when we present results.

The quark propagator allows one, in principle,
to determine the mass renormalization factor $Z_m$,
using
\begin{equation}
\frac1{48}\Tr\left[(\overline{\overline{I\otimes I}}) S^{-1}(p')\right]
= Z_q'(p')\left[Z_m(p') m + C_1 \frac{\langle\bar\chi\chi\rangle}{p^{'2}}
\right]
\,.
\label{eq:OPE}
\end{equation}
Here we display the leading non-perturbative correction,
obtained in Refs.~\cite{Politzer,deRafael} using the operator product expansion.
In practice, as is well known,
this method of determining $Z_m$ has larger non-perturbative corrections
than that (to be described in Sec.~\ref{subsec:vertex}) using vertex functions.

\subsection{Covariant quark bilinears}
\label{subsec:covariant}

Before discussing vertex functions we introduce the
bilinear operators used in our numerical calculations.
The conventional choice for bilinears relies on a partitioning
of the lattice into $2^4$ hypercubes.
For operators at zero momentum, which is all we consider here,
these take the form
\begin{equation}
\CO_{S\otimes F} = \frac{1}{N_y} \sum_y \sum_{A,B} 
\bar\chi_A(y) \singlebar{S}{F}_{AB}\;\CU_{y+A,y+B}\;\chi_B(y)
\,.
\label{eq:Ohpc}
\end{equation}
Here $y$ labels hypercubes as above, with $N_y$ being the
total number in the lattice.
The hypercube fields are defined by~\cite{DK}
\begin{equation}
\bar\chi_A(y) = \frac14\chi(y+A) \ \ {\rm and}\ \
\chi_B(y) = \frac14\chi(y+B)\,.
\end{equation}
The normalization is such that, in the continuum limit,
the matrix element of $\CO_{S\otimes F}$ is the same as that
of $a^3 \int d^4x \;\bar Q (\gamma_S\otimes \xi_F) Q /V$,
with $V$ the four-volume~\cite{PS}.

The bilinears are made gauge invariant by the inclusion
of $\CU_{y+A, y+B}$, which is the average over
products of gauge links along minimal-length paths connecting
the $\bar\chi$ and $\chi$ fields.
We have investigated various choices of links:
\begin{enumerate}
\item
For unimproved or asqtad valence quarks,
a possible choice is the original gauge links, tadpole-improved: $U_\mu/u_0$.
We find that this leads in general to $Z$-factors differing substantially
from unity, and poor convergence of perturbative predictions.
We do not present results for this choice.
\item
For asqtad valence quarks one can also use the Fat7 $+$ Lepage smeared
links $W_\mu$. The resulting links are closer to unity, and
couple less strongly to gluons with momenta of ${\cal O}(1/a)$.
This is the choice for which we present results with asqtad quarks.
\item
For HYP valence quarks we use HYP-smeared links.
\end{enumerate}

The operators (\ref{eq:Ohpc}) do not, in general,
transform irreducibly under
translations, because they rely on a particular partitioning of
the lattice into hypercubes.
As discussed in Ref.~\cite{SP}, they can be
written as linear combinations of operators with definite,
and in general different, transformation properties.
These operators are distinguished by having different
numbers of derivatives and thus varying dimensions.
The operators of lowest dimension are those with no
derivatives and thus $d=3$: these are 
the ``translationally covariant'' (``covariant'' for short)
hypercube operators.

Although non-covariant four-fermion operators
are being used in the calculations of $B_K$ with staggered fermions,
we have chosen to use covariant bilinears in the present study.
This is because these operators are simpler to code, and 
have simpler renormalization properties.
Indeed, if one were calculating matrix elements of staggered bilinears,
such as those needed for $K\to\pi$ semileptonic form factors,
then covariant bilinears would be a natural choice.

The explicit form of these operators was not determined in Ref.~\cite{SP},
so we construct them here.
A simple approach is to adapt the methodology 
developed in Ref.~\cite{GSbaryons} for the construction of
irreducible baryon operators.
The key point is that, when separating the quark
and antiquark fields in the bilinear, one obtains objects
which transform irreducibly under translations if one uses
``symmetric shifts''. These are shifts in which one averages over
forward and backward directions (including, of course, the
gauge links necessary for gauge invariance).
The operator in Eq.~(\ref{eq:Ohpc}) is not of this form.
For example, for a vector current with $S=(1000)$ and $F=(0000)$,
if $A=(0000)$ then $B=(1000)$ and one only has the link pointing
in one direction. A symmetric shift would include terms with
$A=(0000),\;B=(-1000)$ as well as $A=(2000),\;B=(1000)$
(each weighted by a factor of $1/4$) in addition to the original
term (with a weight of $1/2$ since it appears both when shifting
the $\chi$ field and the $\bar\chi$).\footnote{%
Here it is convenient to allow the vectors $A$ and
$B$ to range outside the hypercube.}

This example illustrates the general prescription
for converting the hypercube operators (\ref{eq:Ohpc})
into covariant operators. For given values of $A$ and $B$
(and recalling that, for fixed $S$ and $F$, only one value of
$B$ contributes for each $A$, namely $B=_2 A+S-F$ [with the subscript
indicating mod-2 arithmetic]), one 
replaces $\bar\chi_A\; {\cal U}_{y+A,y+B}\; \chi_B$
with
\begin{align}
\begin{split}
\frac{1}{2 N_{\Delta}} 
\sum_{\Delta}
&\Big(
\bar\chi_A\; {\cal U}_{y+A,y+A+\Delta}\; \chi_{A+\Delta}
\\
&+ 
\bar\chi_{B-\Delta}\; {\cal U}_{y+B-\Delta,y+B}\; \chi_{B}
\Big)\,,
\end{split}
\label{eq:symmsum}
\end{align}
where the set of $N_{\Delta}$ allowed vectors $\Delta$ 
are those obtained from $B-A$ by independently
changing the signs of the non-zero components, including
no changes. For example, if $B-A=(1100)$, then
\begin{equation}
\Delta=(1100),(\!-\!1100),(1\!-\!100),(\!-\!1\!-\!100)\,,
\end{equation}
and so $N_{\Delta}=4$.

After some algebraic manipulations, the resulting operator
can be written 
\begin{eqnarray}
\lefteqn{\CO_{S\otimes F}^{cov} = 
\frac{1}{N_{\rm site}} \sum_n \frac{1}{16} \sum_{A,B} }
\nonumber\\
&&\ \ \bar\chi(n) \singlebar{S}{F}_{n,n\!+\!B\!-\!A} \;
\CU_{n,n\!+\!B\!-\!A}\; \chi(n\!+\!B\!-\!A)
\,.
\label{eq:Ocov}
\end{eqnarray}
The factor of $1/16$ is required in order to retain the
same normalization as in (\ref{eq:Ohpc}), because of
the definition $\chi(y)_A= (1/4) \chi(y+A)$.
The double sum over $A$ and $B$ in (\ref{eq:Ocov}),
which is really a single sum since
$\singlebar{S}{F}$ enforces $B=_2 A+S+F$,
corresponds to the sum over $\Delta$ in the symmetric shift.
This can be made explicit by writing the operator as
\begin{eqnarray}
\lefteqn{
\CO_{S\otimes F}^{cov} = 
\frac{1}{N_{\rm site}} \sum_n 
\frac{1}{N_\Delta} \sum_{|\Delta|=|S\!-\!F|} }
\nonumber\\
&&\ \ 
\bar\chi(n) \singlebar{S}{F}_{n,n\!+\!S\!-\!F} \;
\CU_{n,n\!+\!\Delta}\; \chi(n\!+\!\Delta)
\,,
\label{eq:Ocovb}
\end{eqnarray}
where the second sum is over the $N_\Delta$ allowed
values of $\Delta$.
This result makes the presence of symmetric shifts manifest.
Note that the sign arising from $\singlebar{S}{F}$ is independent
of $\Delta$, and that the form (\ref{eq:Ocovb}) removes
some redundancy in the sums of (\ref{eq:Ocov}).

The forms (\ref{eq:Ocov}) and (\ref{eq:Ocovb}) show
explicitly that the covariant bilinears do not require a
partitioning of the lattice into hypercubes.
This simplifies their numerical implementation, since one
can freely sum over $n$.
The computation of the link factors is the most costly
part of the calculation, 
with the cost growing rapidly with $|\Delta|$.

\subsection{Vertex renormalization}
\label{subsec:vertex}

To determine matching factors of general bilinears we must
calculate the vertex functions. We consider here only
the case of exceptional kinematics in which the operator
inserts no momentum:
\begin{equation}
\Lambda_{AB}^{S\otimes F}(p')
=
\frac1{N_{\rm site}}
\langle \phi_A^{a} (p')\; \CO_{S\otimes F}^{cov;\; (ab)}\; \bar\phi^{b}_B(-p')
\rangle
\,.
\label{eq:Lambdadef}
\end{equation}
Like the propagator, the vertex is $48\times48$ matrix,
with the color part being trivial.
The new indices $a$ and $b$ in the superscripts are flavor indices.
We always choose $a\ne b$ so that the operator is a flavor non-singlet,
which implies that there is only a single quark contraction between
the external fields and the operator.
The fields in the vertex are valence quarks and antiquarks,
as for the propagator.

One now follows the perturbative renormalization procedure,
amputating the vertex with the previously calculated
inverse propagators:
\begin{equation}
\Gamma^{S\otimes F}(p') = S^{-1}(p') \Lambda^{S\otimes F}(p')
S^{-1}(p')\,.
\label{eq:Gammadef}
\end{equation}
Matching factors are determined by enforcing the
tree-level form of $\Gamma$ when fields and operators are
renormalized:
\begin{equation}
\frac{Z'_q(p')}{Z_{S\otimes F}(p')}
=
\frac1{48}
\frac{\Tr\left[\doublebar{S}{F}^\dagger \Gamma^{S\otimes F}(p')\right]}
     {V_{S\otimes F}(p')}
\,.
\label{eq:NPRvertex}
\end{equation}
Here we assume no mixing, which is the case for the covariant bilinears.
This is shown non-perturbatively in App.~\ref{app:symm}.
We have also divided the projected vertex by its tree-level expression,
$V_{S\otimes F}$. This has the form $1+O(a^2)$,
and is given explicitly in Eq.~(\ref{eq:vertexfunction}).
Dividing by $V$  removes some of the discretization errors, and
this approach is common practice in NPR.

One can use
the following lattice Ward identities to relate matching factors:
\begin{eqnarray}
\frac1m\Tr\left[\overline{\overline{(I\otimes I)}}\; S^{-1}(p)\right]
&=&
\Tr\left[\doublebar55\; \Gamma^{5\otimes 5}(p)\right]\,,
\label{eq:WIP}
\\
\frac{\partial}{\partial m}
\Tr\left[\overline{\overline{(I\otimes I)}}\; S^{-1}(p)\right]
&=&
\Tr\left[\overline{\overline{(I\otimes I)}}\; \Gamma^{I\otimes I}(p)\right]
\,,
\label{eq:WIS}
\end{eqnarray}
These follow by standard manipulations,
and hold as written only when $m$ is the valence quark mass 
[so that the derivative in (\ref{eq:WIS}) does not act on sea quark masses]
and the operators in the vertices are flavor non-singlets
[so that there are no ``quark-disconnected'' contractions].
Using the definition (\ref{eq:NPRvertex}) for the right-hand sides
and inserting the result (\ref{eq:OPE}) into the left-hand sides, we find,
at sufficiently large ${p'}^2$, that
\begin{eqnarray}
\lefteqn{Z_q(p') Z_m(p') = \frac{Z_q(p')}{Z_P(p')}=\frac{Z_q(p')}{Z_S(p')}}
\nonumber\\
&&
\Rightarrow Z_m(p') = \frac1{Z_P(p')} = \frac1{Z_S(p')}\,.
\label{eq:WIres}
\end{eqnarray}
These are the familiar relations from continuum perturbation theory,
which here hold non-perturbatively.

We can now see why it is better to use the vertex rather than
the propagator [Eq.~(\ref{eq:OPE})] to determine $Z_S=1/Z_m$.
This is because the condensate term in (\ref{eq:OPE}),
which gives a significant correction at typical values of $p'$,
is absent in the scalar vertex.
This can be seen by
inserting (\ref{eq:OPE}) in the left-hand side of (\ref{eq:WIS}).
The condensate appearing in the operator product expansion is evaluated
in the chiral limit, so the $\partial/\partial m$ removes this $1/{p'}^2$
contribution.
By contrast, a similar analysis for the pseudoscalar vertex
shows that there is a non-perturbative correction 
proportional to $\langle\bar q q\rangle/(m {p'}^2)$.
This is the well-known pion pole contribution~\cite{NPR}, which
makes the direct determination of $Z_P$ difficult.

One can also use axial Ward identities to show that
$Z_{S\otimes F}=Z_{S5\otimes F5}$, where 
the subscript $S5\otimes F5$ indicates the bilinear with
spin-taste $\overline{(\gamma_S\gamma_5\otimes \xi_F\xi_5)}$.
We do not reproduce the derivation as this result
is already known to hold to all orders in perturbation theory~\cite{SP}.

\subsection{Irreducible representations and perturbative matching}
\label{subsec;irreps}

\begin{ruledtabular}
\begin{table*}[htb!]
\begin{center}
\begin{tabular}{c | c | c | c}
\# links & S & V & T \\ \hline 
4 
&$(I \otimes \xi_5)$
&$(\gamma_\mu \otimes \xi_\mu \xi_5)$
&\,\,$(\gamma_\mu \gamma_\nu \otimes \xi_\mu \xi_\nu \xi_5)$
\\ 
3 
&$(I \otimes \xi_\mu \xi_5)$
&
 $(\gamma_\mu \otimes \xi_{5}) \quad\, 
  (\gamma_\mu \otimes \xi_\nu \xi_\rho)$
&
 $[(\gamma_\mu \gamma_\nu \otimes \xi_\mu \xi_5) 
  \quad  (\gamma_\mu \gamma_\nu \otimes \xi_\rho)]$ 
\\ 
2
&$(I \otimes \xi_\mu \xi_\nu)$
&$(\gamma_\mu \otimes \xi_\nu) \quad 
  (\gamma_\mu \otimes \xi_\nu \xi_5)$ 
&$[(\gamma_\mu \gamma_\nu \otimes I)
  \quad \quad \,\,\, 
    (\gamma_\mu \gamma_\nu \otimes \xi_5)] 
  \quad (\gamma_\mu \gamma_\nu \otimes \xi_\nu \xi_\rho)$ 
\\ 
1
&$(I \otimes \xi_\mu)$
&$(\gamma_\mu \otimes I)
  \quad\,\,\,\, (\gamma_\mu \otimes \xi_\mu \xi_\nu)$ 
&$[(\gamma_\mu \gamma_\nu \otimes \xi_\nu) 
   \quad  (\gamma_\mu \gamma_\nu \otimes \xi_\rho \xi_5)]$ 
\\
0
&$(I \otimes I)$
&$(\gamma_\mu \otimes \xi_\mu)$
&\,\,$(\gamma_\mu \gamma_\nu \otimes \xi_\mu \xi_\nu)$
\end{tabular}
\end{center}
\caption{Spin-taste assignments
of covariant bilinears forming irreps of the lattice symmetry group.
Indices $\mu$, $\nu$ and $\rho$ are summed from $1-4$, except
that all are different. 
If two indices appear in either the spin or the taste, there
may be some redundancy, e.g.\ in $(1\otimes\xi_\mu\xi_\nu)$ one
can enforce $\mu<\nu$ so that the dimension of the irrep is 6.
Pseudoscalar and axial bilinears are not listed:
they can be obtained from scalar and vector, respectively, by
multiplication by $\gamma_5\otimes\xi_5$. Bilinears related
in this way have the same matching factors. This operation also implies
the equality of the matching factors for the three pairs of tensor bilinears
within square brackets.
}
\label{tab:irreps}
\end{table*}
\end{ruledtabular}

The $16^2$ covariant bilinears $\CO_{S\otimes F}^{\rm cov}$
fall into 35 irreps under the lattice symmetry group.
These are collected in Table~\ref{tab:irreps}, organized
according to the number of links, i.e.\ the separation between
quark and antiquark fields. To our knowledge, this decomposition
into irreps for covariant bilinears has not been demonstrated
previously in the literature. Thus we provide a brief
demonstration in App.~\ref{app:irreps}.

As already noted above, matching factors for operators with
spin-taste $(\gamma_S\otimes \xi_F)$ and
$(\gamma_S\gamma_5\otimes \xi_F\xi_5)$ are the same.
This reduces the number of independent matching factors from 35 to
19, as described in the caption to the table.

Since our aim is to compare to perturbation theory,
we need the one-loop matching factors for the covariant operators.
It turns out, for reasons discussed in App.~\ref{app:pert}, that
they can be obtained from those for hypercube bilinears in
a trivial way: one simply has to drop the mixing terms,
with the diagonal matching factors being unchanged.
The lack of mixing is a direct result of using
covariant operators, since different spin-tastes lie in
different irreps of the lattice symmetry group.

Expressions for the required diagonal matching factors are given
in Ref.~\cite{KLS} in terms of a single lattice loop integral.
Numerical values are, however, not given for the HYP-smearing
coefficients that we use, nor for mean-field improved asqtad bilinears.
We have calculated these values and collect them in App.~\ref{app:pert}.

\section{Numerical implementation}
\label{sec:numerics}

We use the Chroma~\cite{Edwards:2004sx} software library for
Landau-gauge fixing, HYP smearing, 
and asqtad inversions.  
We have added code to implement momentum sources, 
to invert the unimproved staggered fermion matrix
(needed for HYP-smeared fermions), 
and to construct the bilinears including the gauge links.
Stopping criteria for gauge fixing and propagator inversions 
were set so that the errors are smaller than those from other sources,
and in particular from statistics~\cite{Lytlethesis}.

Our gauge configurations are taken from the MILC coarse ($a \approx
0.12 \text{ fm}$) and fine ($a \approx 0.09 \text{ fm}$) 
ensembles~\cite{milcrmp}, which are generated using
asqtad fermions and Symanzik-improved gauge action.
Relevant details are given in Tables~\ref{tab:ASQcoarse} and \ref{tab:ASQfine}.
We include results for
the $\tilde u_0$ factors needed for mean-field improvement;
these are defined in App.~\ref{app:pert}.

\begin{ruledtabular}
\begin{table*}[htb!]
\begin{center}
\begin{tabular}{c | c | c | c | c | c | c}
$am_{\text{sea}}$/$am_{\text{val}}$ & \# configs & \# momenta & $a^{-1} \,\,[\text{GeV}]$ & $u_0$ & $\tilde u_0^{\rm HYP}$ & $\tilde u_0^{\rm ASQ}$ \\ \hline 
 0.03/0.03 & 16 & 7/10 & 1.682 & 0.8696 & 0.9845 & 1.0521 \\
 0.02/0.02 & 16 & 7/10 & 1.679 & 0.8688 & 0.9843 & 1.0525\\
 0.01/0.01 & 16 & 7/10 & 1.662 & 0.8677 & 0.9841 & 1.0528\\
\text{chiral} & - & 7/10 & 1.654 & -      & 0.9839 & 1.0532
\end{tabular}
\end{center}
\caption{Parameters of coarse ensembles. 
The lattices are of size $20^3\times 64$.
The quoted masses are for the light (average of up and down)
quarks, there is in addition a strange sea quark of fixed bare
mass $am_{\rm sea,strange}=0.05$.
Lattice spacings are obtained using
$r_1=0.3108\;$fm and taken from Ref.~\cite{milcrmp}.
Extrapolations to the chiral limit are 
done with a linear fit. The quoted number of momenta are for valence 
asqtad/HYP-smeared fermions.}
\label{tab:ASQcoarse}
\end{table*}
\end{ruledtabular}

\begin{ruledtabular}
\begin{table*}[htb!]
\begin{center}
\begin{tabular}{c | c | c | c | c | c | c}
$am_{\text{sea}}$/$am_{\text{val}}$ & \# configs & \# momenta & $a^{-1} \,\,[\text{GeV}]$ & $u_{0}$ & $\tilde u_0^{\rm HYP}$ & $\tilde u_0^{\rm ASQ}$ \\ \hline 
0.0124/0.0124 & 16 & 8 &  2.357 & 0.8788 & 0.9869 & 1.0507\\
0.0093/0.0093 & 16 & 8 &  2.352 & 0.8785 & 0.9868 & 1.0508\\
0.0062/0.0062 & 16 & 8 &  2.349 & 0.8782 & 0.9868 & 1.0509\\
\text{chiral} & -  & 8 &  2.340 & -      & 0.9867 & 1.0511
\end{tabular}
\end{center}
\caption{Parameters of asqtad fine ensembles. Lattices are of size
$28^3\times 96$, and the strange sea quark mass is 
$am_{\rm sea,strange}=0.031$.}
\label{tab:ASQfine}
\end{table*}
\end{ruledtabular}

The momenta $p'$ that we use are listed in Table~\ref{tab:momenta}. 
These are chosen so that the components
are comparable in all four directions
(after inclusion of $2\pi/L_{s,t}$ factors), 
ensuring that no single component becomes too large. 
This is known to reduce discretization errors.
These choices cover the expected NPR window, as will be seen below.

\begin{ruledtabular}
\begin{table*}[htb!]
\begin{center}
\begin{tabular}{c | c | c }
lattice & fermion & momenta  \\ \hline 
coarse & asqtad & (1,2,2,4), (2,1,2,6), (2,2,2,7), (2,2,2,8), (2,2,2,9), (2,3,2,7), (3,3,3,9)\\
fine & asqtad & (1,2,2,5), (2,2,2,6), (2,2,2,7), (2,2,2,8), (2,2,3,8), (2,3,3,9), (3,3,3,10), (3,3,3,12)\\
coarse & HYP & (1,1,1,4), (1,1,1,6), (1,2,1,5), (1,2,2,4), (2,1,2,6), (2,2,2,7), (2,2,2,8), (2,2,2,9), (2,3,2,7), (3,3,3,9)\\
fine & HYP & (1,2,2,5), (2,2,2,6), (2,2,2,7), (2,2,2,8), (2,2,3,8), (2,3,3,9), (3,3,3,10), (3,3,3,12)
\end{tabular}
\end{center}
\caption{Physical momenta used in our calculations.
The four vectors are in units of $(2\pi/L_s,2\pi/L_s,2\pi/L_s,2\pi/L_t)$,
where $L_s$ ($L_t$) is the number of 
sites in the spatial (temporal) directions.}
\label{tab:momenta}
\end{table*}
\end{ruledtabular}

For every gauge-fixed configuration in our ensemble $\{ U^{i} \}$
and each physical momentum $p'$ under consideration, we invert the
Dirac operator $D$ on 16 momentum sources, 
solving $DS = e^{ip \cdot n}$ for
$p=p' + \pi B$ to obtain 
\begin{equation}
S^{i}(n, p'+\pi B) = \langle \chi(n) \bar\phi_B(-p')\rangle_{U^i}
\,,
\end{equation}
where color indices are suppressed.
We next Fourier
transform the free space index with the 16 different momenta
$p=-p'+\pi A$, leading
to the $16\times 16$ momentum-space propagator matrix
$S^{i}(p')_{AB}$ of Eq.~(\ref{eq:Sdef}).
This is then averaged over configurations to obtain the
propagator $S(p')_{AB}$ of~\eqref{eq:Sdef}.
Lattice symmetries predict that the inverse propagator contains
only $1\otimes1$ and $\gamma_\mu\otimes1$ contributions
up to terms suppressed by $a^4$
[see Eq.~(\ref{eq:Sinvform})]. We have checked that
non-continuum terms are in fact consistent with zero within our 
statistical errors.

Vertex functions are constructed from $S^{i}(n, p'+\pi B)$
and
\begin{equation}
S^{i}(p'+\pi A, n) = 
\langle \phi_A(p') \bar\chi(n)\rangle_{U^i}\,.
\end{equation}
As usual, the latter propagator may be obtained from the former
using the staggered analogue of $\gamma_{5}$-hermiticity of the Dirac
operator, $\epsilon D^{\dagger} \epsilon = D$, where 
$\epsilon=(-1)^{n_1+n_2+n_3+n_4} \equiv (-1)^n$ 
is the alternating phase factor.
We find
\begin{equation}
S^{i}(p' + \pi A, n) = (-1)^{n} S^{i}(n, p' + \pi\tilde{A})^{\dagger}\,,
\end{equation}
where $\tilde{A} = A +_{2} (1,1,1,1)$, and the
hermitian conjugation acts on color indices.
The two propagators are then tied together with the bilinear.
For example, the pseudoscalar (unamputated) vertex is
\begin{align}
&
\Lambda_{AB}^{\gamma_5\otimes \xi_5}(p') =
\langle \phi_A (p') \CO_{\gamma_5 \otimes \xi_5}^{cov} \bar\phi_B(-p')
\rangle 
\\ 
&= \frac{1}{N_{\text{conf}} N_{\rm site}} \sum_{i,n} 
S^{i}(p'\! +\! \pi A, n)(-1)^n S^{i}(n, p'\!+\!\pi B) \nonumber 
\\ 
&= \frac{1}{N_{\text{conf}} N_{\rm site}} 
\sum_{i,n} S^{i}(n, p'\!+\!\pi \tilde A)^{\dagger} S^{i}(n, p'\!+\!\pi B) 
\,. \nonumber
\end{align}
In the general the two
propagators end at positions differing by a hypercube vector,
and are connected by an average over products of links
over minimal-length paths [cf.\ Eq.~(\ref{eq:Ocovb})].

Amputation and determination of the $Z$-factors is then performed
using Eqs.~(\ref{eq:Gammadef}) and (\ref{eq:NPRvertex}).
For a given value of $p'$, these involve manipulations
of $16\times 16$ matrices, which can be done in the analysis phase
of the calculation.
The $Z$-factor corresponding to a given irrep is determined by averaging
the traces of the amputated vertex functions in that irrep; e.g.\ for the vector
we compute $Z_V$ from
$\frac{1}{4}\Tr \sum_\mu  (\overline{\overline{\gamma_\mu \otimes 1}} )\, \Gamma^{\gamma_\mu \otimes 1}(p')$.

\begin{figure}[tb]
\begin{center}
\includegraphics[]{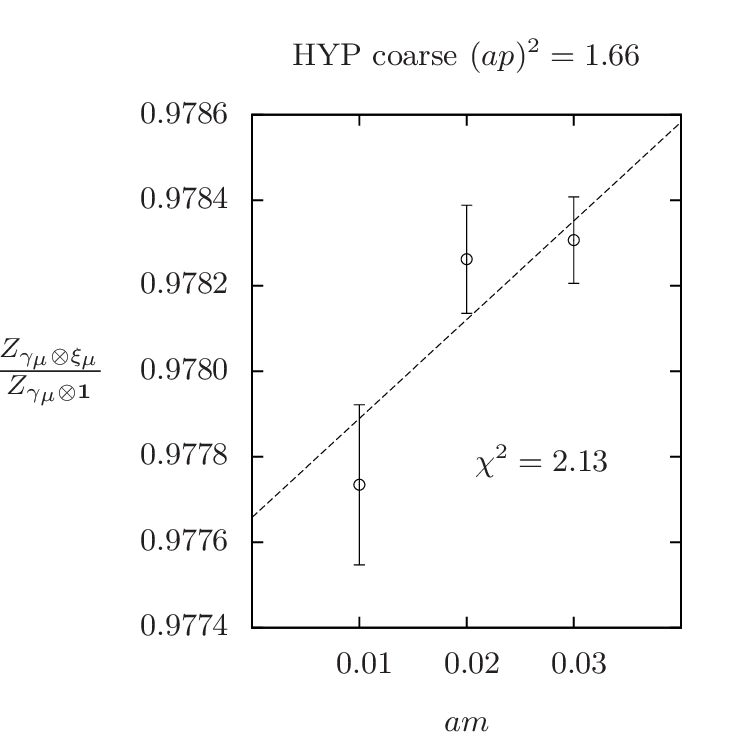}
\caption{Example of chiral extrapolation for the ratio
$Z_{\gamma_\mu\otimes \xi_\mu}/Z_{\gamma_\mu\otimes 1}$ on the HYP coarse 
ensemble.}
\label{fig:coarse_chiral_vector1}
\end{center}
\end{figure}

\begin{figure}[tb!]
\begin{center}
\includegraphics[]{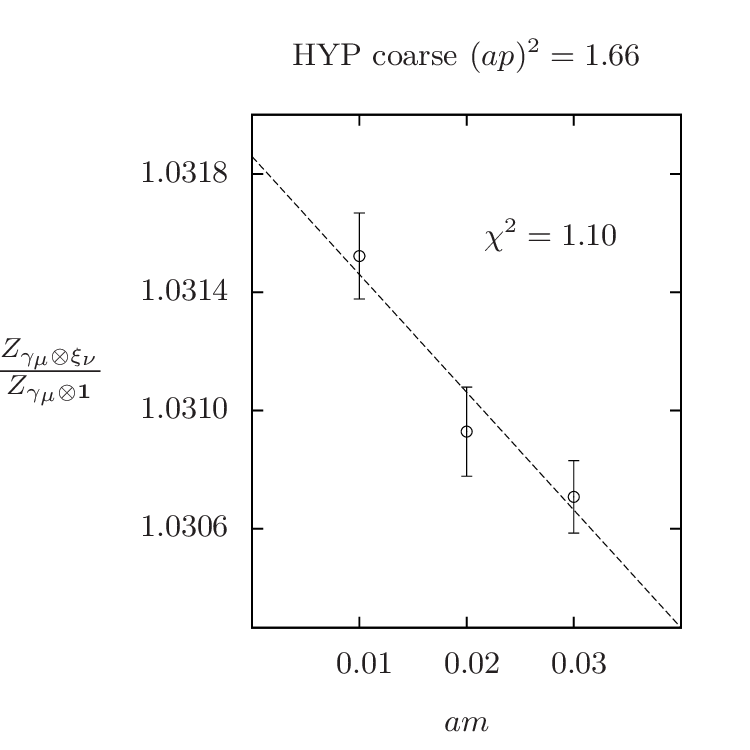}
\caption{Example of chiral extrapolation for the ratio
$Z_{\gamma_\mu\otimes \xi_\nu}/Z_{\gamma_\mu\otimes 1}$ on the HYP coarse ensemble.}
\label{fig:coarse_chiral_vector3}
\end{center}
\end{figure}

For both asqtad and HYP-smeared fermions, and for both coarse and
fine lattices, we use valence quarks with bare masses equal to those
of the light (asqtad) sea quarks. Thus our calculations are unquenched for
asqtad valence fermions. For HYP-smeared fermions we are, however,
using a mixed fermion action (different discretizations of valence and
sea quarks) and, in addition, a partially quenched set-up
(because $Z$-factors for HYP-smeared and asqtad
fermions are different, so that
the physical masses of sea and valence quarks differ
even though the bare masses are equal).
For both types of valence fermions 
we extrapolate our final results to the chiral limit
using a linear fit. If the dependence on quark masses is linear and
weak, this extrapolation will remove partial quenching effects
for the HYP-smeared fermions. Residual effects from using a mixed action
should vanish in the continuum limit, and thus
appear as additional discretization errors for $a>0$.
Examples of the chiral fits for $Z_{\gamma_\mu\otimes 1}$
are shown in Figs.~\ref{fig:coarse_chiral_vector1}
and~\ref{fig:coarse_chiral_vector3}.
These are typical in terms of the quality of fits, 
although the extent of the chiral extrapolation is greater
for scalar bilinears.
We also use linear chiral fits to determine values for
$1/a$, $\tilde u_0^{\rm HYP}$ and $\tilde u_0^{\rm ASQ}$ which
we use in subsequent analysis.
These are shown in Tables~\ref{tab:ASQcoarse} and \ref{tab:ASQfine}.
We stress that these are very
mild extrapolations, so that none of our conclusions
would be changed were we to take the values of these
quantities from, say, the lattice spacing with the smallest
values of the valence quark masses.

The only exception to the above discussion of chiral extrapolations
are the matching factors for pseudoscalar bilinears.
As discussed after Eq.~(\ref{eq:WIres}),
these are singular in the $m \rightarrow 0$ limit~\cite{NPR}.
It is possible to remove the singular part in various ways,
but in this work we have chosen to exclude the
pseudoscalars from our analysis. 

Although we can extrapolate to the chiral limit for the
two light quarks, our calculations have the
strange sea-quark mass fixed at approximately its physical value.
Strictly speaking, this means that our NPR results are not
in the desired mass-independent renormalization scheme.
However, given the mild dependence on quark mass that we observe,
we expect that this shortcoming will have little impact on the
final results. In particular, we assume the error that this
introduces to be smaller than the truncation errors in the
one-loop PT expressions to which we compare.

We compute diagonal $Z$-factors for all 256 choices of 
spin and taste. We have checked in some cases that off-diagonal
contributions to $Z$-factors are consistent with zero, 
as expected given that the covariant bilinears do not mix.
We use 16 decorrelated configurations, which we find to be
sufficient when using momentum sources.
We then combine the 256 choices into the irreps listed in
Table~\ref{tab:irreps}, which further reduces the errors.
All errors are obtained using single-elimination jacknife.

\section{Results}
\label{sec:results}

We divide our discussion of the results into three parts.
In the first two we consider ratios of $Z$-factors in which the
numerator and denominator have the same spins but different tastes.
Specifically we consider ratios in which the denominators are
taste singlets:\footnote{%
In this section we will denote the NPR momentum scale by $p$, 
which has physical units. Thus $a p$ here corresponds to the $p'$
used in previous sections. 
We also make explicit that the $Z$-factors depend separately
on $p$ and $a$ in general.}
\begin{eqnarray}
\lefteqn{\frac{Z_{S\otimes F}(p,a)}{Z_{S\otimes 1}(p,a)} =}
\nonumber\\
&& 1 +
\frac{\alpha(\mu_0)}{4\pi} \left[ C^{\rm LAT}_{S\otimes F} 
\!-\! C^{\rm LAT}_{S\otimes 1} \right]
+ {\cal O}([ap]^2) + \dots  \,.
\label{eq:rationoMFI}
\end{eqnarray}
As discussed in App.~\ref{app:pert},
PT predicts these ratios to be independent of the
NPR momentum $p$ since they are dominated by contributions
from loop momenta near the cut-off scale.
In particular, $\alpha$ is to be evaluated
at a scale $\mu_0\sim 1/a$ which is not related to $|p|$. 
This is illustrated by the right-hand side 
of Eq.~(\ref{eq:rationoMFI}), which shows the one-loop expression
for the simplified case of no mean-field improvement.\footnote{%
The $C^{\rm LAT}$ are finite lattice constants. 
For the general expression including
mean-field improvement see Eqs.~\eqref{eq:Zmaster}--\eqref{eq:deltadef}.}
These ratios are thus good quantities to use to test the accuracy of
PT since one does not have to worry about anomalous dimensions.
They are analogous to $Z_A/Z_V$ with Wilson-like fermions, with
the analogue of the lack of $p$ dependence being the fact that
$Z_A/Z_V$ calculated in different ways should agree up to discretization
errors.

The lack of dependence of the ratios on $p$
does not carry over to discretization effects
[represented in Eq.(\ref{eq:rationoMFI}) by the $(ap)^2$ term]
or to non-perturbative effects, which behave as
inverse powers of $p$ and are important only for small $p$. 
One can hope to disentangle these
effects by studying the $p$ and $a$ dependence of the ratios, 
as we discuss below.

In the final part of this section we present results for the
denominators of the ratios. These do have anomalous dimensions,
so we can see how well the $p$ dependence agrees with the perturbative
predictions. These predictions can be made using continuum
perturbation theory, for which results are known to three or four loop order
(as described in App.~\ref{app:running}).

\subsection{Ratios for HYP-smeared bilinears}

\begin{figure*}[tb!]
\begin{center}
\includegraphics[]{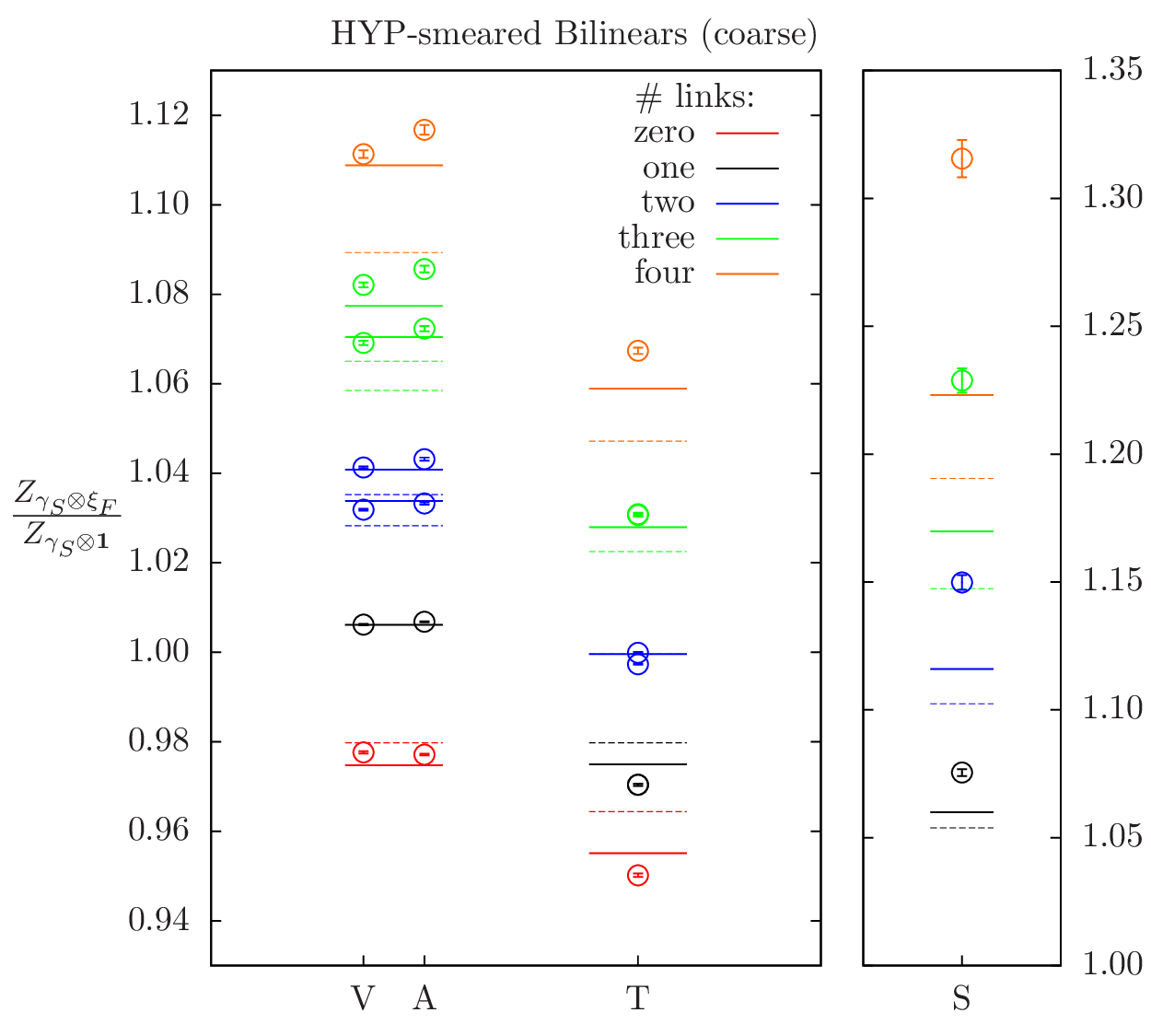} 
\caption{Comparison of $Z$-factor ratios obtained using NPR
to one-loop perturbation theory
for HYP fermions on coarse attices.  
V, A, T and S refer to bilinears with vector, axial, tensor and
scalar spins, respectively.
Horizontal lines show perturbative predictions, with
solid/dotted lines showing results with/without mean-field improvement.
Results are in the chiral limit for the momentum described in the
text.
}
\label{fig:HYP_ratios_both}
\end{center}
\end{figure*}

\begin{figure*}[tb!]
\begin{center}
\includegraphics[]{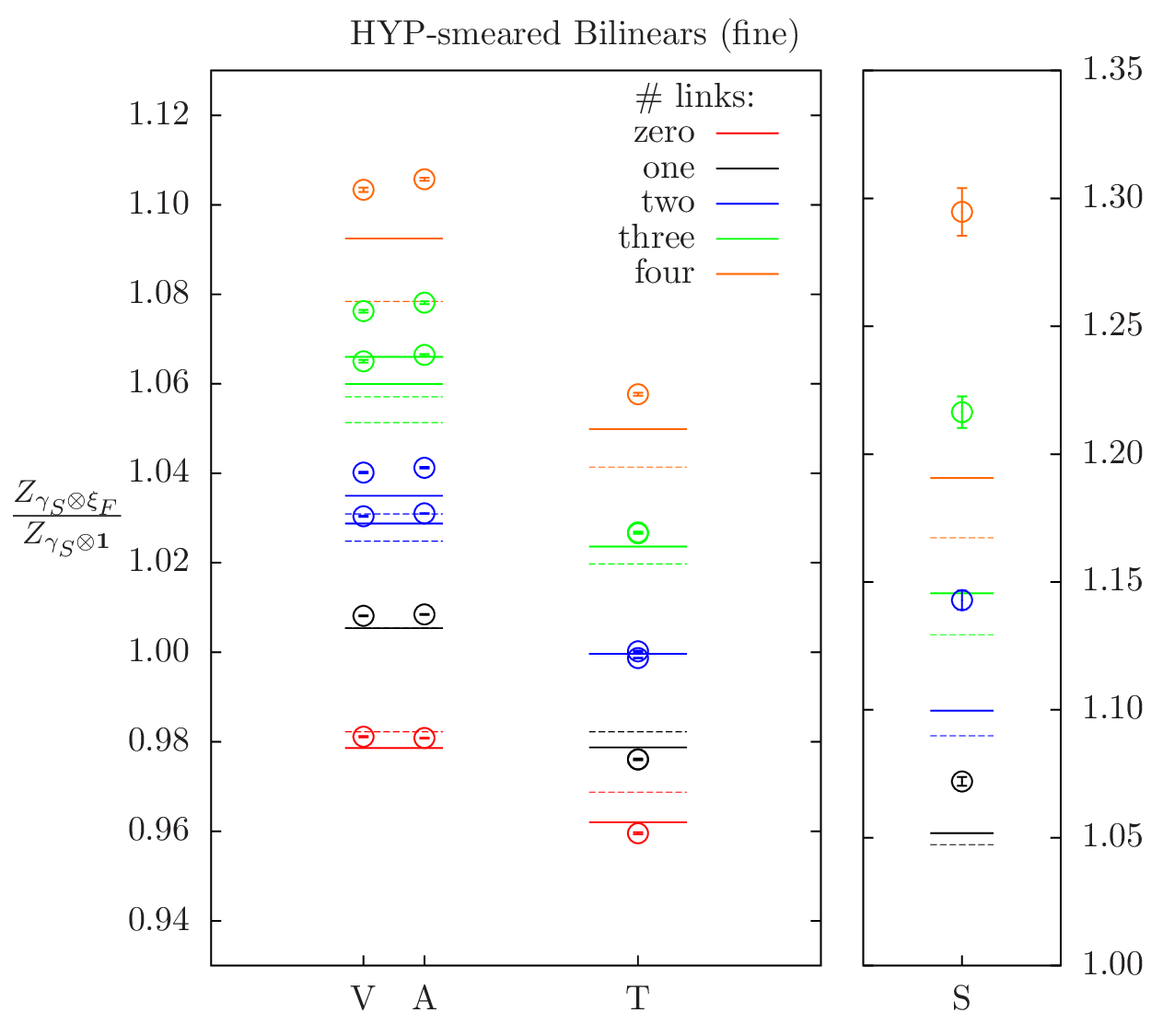}
\caption{As for Fig.~\ref{fig:HYP_ratios_both} but on the fine
lattices.}
\label{fig:HYP_fine_ratios_both}
\end{center}
\end{figure*}

We begin by discussing the results with HYP-smeared fermions.
In Figs.~\ref{fig:HYP_ratios_both} and \ref{fig:HYP_fine_ratios_both}
we display results for all ratios at a fixed NPR momentum.
We choose $a p =(2,2,2,7)$ 
in units of $(2\pi/L_s,2\pi/L_s,2\pi/L_s,2\pi/L_t)$,
so that $(ap)^2\approx 1.66$ and $0.81$, respectively,
on coarse and fine lattices.
This turns out to correspond to nearly the same physical value,
$|p| \approx 2.1\;$GeV, for both lattice spacings.
We expect that this choice satisfies the
window condition~\eqref{eq:npr_condition} for both lattice spacings.

These figures show the comparison of the 26 ratios
involving bilinears with vector, axial, tensor and scalar spins
to one-loop PT.
We show perturbative predictions both without
[Eq.~(\ref{eq:rationoMFI})] and with
[Eq.~(\ref{eq:Zratsrelation})] mean-field improvement~\cite{LM}.\footnote{%
We stress that we are using mean-field improvement to obtain
an improved perturbative prediction for the {\em same} operators.}
For these predictions we use $\mu_0=1.8/a$
leading to $\alpha(\mu_0)\approx 0.24$
and $0.21$ on the coarse and fine lattices, respectively.
For the mean-field improved prediction, we also need values for
$\tilde u_0^{\rm HYP}$, which are given in Table~\ref{tab:ASQcoarse}.
The color coding in the plots indicates the number of links 
in the operators in the numerators of the ratios.
The denominators have 1-link operators
for spins V and A, 2-link operators for T, and 0-link operators for S.

Overall the one-loop prediction works well.
We highlight certain features.
First, the statistical errors in the NPR results 
are very small, particularly for spins V, A and T.
Second, PT correctly captures the ordering with link number,
and placement relative to unity.
This ordering is the dominant feature of the results, and
indicates that the fluctuations in individual smeared links
(which reduce their [gauge-fixed] average values below unity)
are the largest contributor to $Z$-factors differing from unity.
Bilinears with more links thus have smaller matrix elements and
require larger $Z$-factors to attain the canonical normalization.
This argument would imply that ratios involving 1-link  
V and A ratios and  2-link T ratios should lie close to unity,
since the numerators and denominators have the same number of links.
This is indeed what is observed.

Third, PT correctly predicts the ``fine structure'' within a given link
number. For example, for spins V and A, 
there are two ratios involving three-link numerators, and
two involving two-links (see Table~\ref{tab:irreps}).
The predicted orderings and splittings match well with PT.
There is a similar fine structure for the tensors, though this
is hard to discern from the figure. For one-link numerators,
there are two ratios, which are predicted to be equal to all orders in
PT. The NPR results for these two ratios are indistinguishable.
The same is true for three-link numerators.
For two-link numerators there are also two ratios, but in this case
they are predicted to be equal at one-loop order but not to all orders.
Here the NPR results for the two ratios do differ, 
but the difference is very small
(and consistent with a two-loop or higher-order perturbative effect).

Fourth, we recall that matching factors for spins V and A
are predicted to be equal to all-orders in PT.
We observe very small (subpercent level) differences.
Differences can arise due to long-distance non-perturbative effects,
and so these effects are small in these channels. 

Fifth, we note that the NPR results
on the fine lattices are all slightly closer to unity 
than those on the coarse lattices. 
This is qualitatively what one would expect if the dominant
contribution to the difference from unity was perturbative,
since $\alpha(\mu_0)$ decreases with $a$ if $\mu_0\sim1/a$.
However, a complete interpretation of this result requires
understanding the contributions of non-perturbative and
discretization errors, which we discuss below.

We now discuss the level of quantitative agreement between the
NPR results and one-loop PT. With the couplings we have chosen,
the agreement is at the subpercent level for spins V, A and
T, and at the 5-10\% level for scalars.
We cannot, of course, expect perfect agreement because we
have truncated PT. One way of estimating the
uncertainty in the one-loop prediction is to vary the
scale at which $\alpha$ is evaluated over a reasonable range.
Were we to use $\mu_0=1/a$ rather than $1.8/a$, the
couplings would become roughly 30\% larger
($\alpha\approx 0.32$ and $0.27$
on the coarse and fine lattices, respectively).
This would lead to a much improved quantitative agreement
with the scalar ratios,
while that with spins V, A and T would be less good.\footnote{%
It would be interesting to use an
approximate scale-setting method to better predict the appropriate
value of $\alpha$ to use for each quantity. Our data suggests
that a lower scale would be found for scalars than for the
other bilinears.}
But the most important point is that, within this perturbative
uncertainty the PT and NPR results agree.

As can be seen from the numerical values,
the coefficients of $\alpha$ in the one-loop predictions
have magnitudes smaller than unity
for all except the three and four-link scalar ratios.
Our way of estimating the uncertainties in
perturbative predictions assumes that a small
one-loop coefficient implies that higher orders are also small.
An alternate, and more conservative, approach is to say that, 
for all ratios, two-loop effects are of size
${\cal O}(1)\times \alpha^2\approx 0.05-0.09$.
This gives a larger estimate than that obtained above
except for the three- and four-link scalar ratios, 
for which the two estimates agree.

Our final comment on these two figures is that we find
the impact of mean-field improvement to be fairly minor,
below the level of the uncertainty due to the choice of $\alpha$.
The effects are small because $\tilde u_0^{\rm HYP}$ is very
close to unity, and lies close to its perturbative prediction
[Eq.~(\ref{eq:u0HYPinPT})].
Evaluating $\alpha$ at the scale $1/a$ the prediction for the
coarse and fine lattices are $0.985$ and $0.988$, respectively,
to be compared to $0.984$ and $0.987$ 
(from Tables~\ref{tab:ASQcoarse} and \ref{tab:ASQfine}).

\begin{figure}[tbp!]
\begin{center}
\includegraphics[]{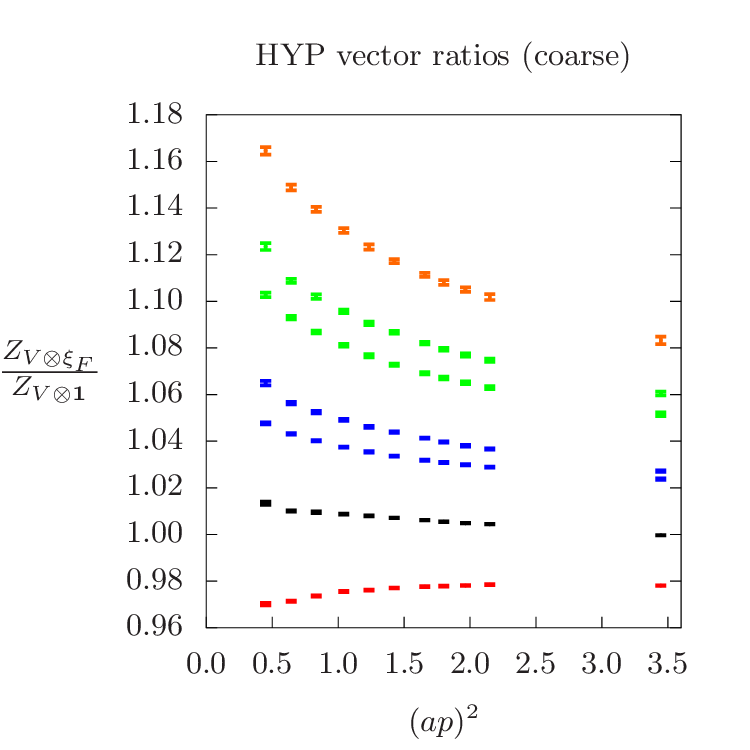}
\caption{Scale dependence of the ratios of vector HYP-smeared
bilinear $Z$-factors on coarse MILC lattices.
The color coding indicates the link-number of the numerator
and corresponds to that in Fig.~\protect\ref{fig:HYP_ratios_both} 
(where the results for $(ap)^2=1.66$ are shown).}
\label{fig:HYP_coarse_ratios_vectors}
\end{center}
\end{figure}

\begin{figure}[tb!]
\begin{center}
\includegraphics[]{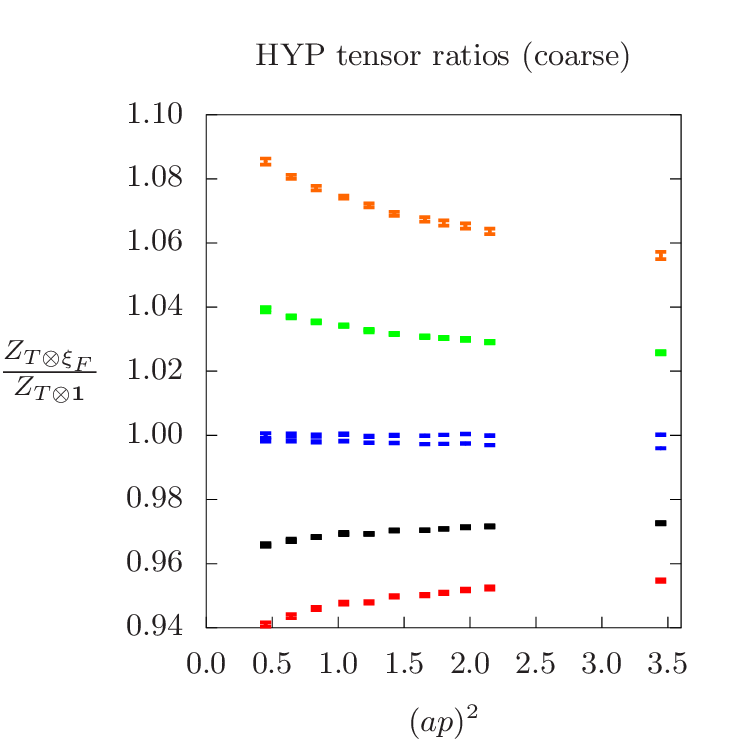}
\caption{As for Fig.~\ref{fig:HYP_coarse_ratios_vectors} but
for the tensor ratios.}
\label{fig:HYP_coarse_ratios_tensors}
\end{center}
\end{figure}

\begin{figure}[tb!]
\begin{center}
\includegraphics[]{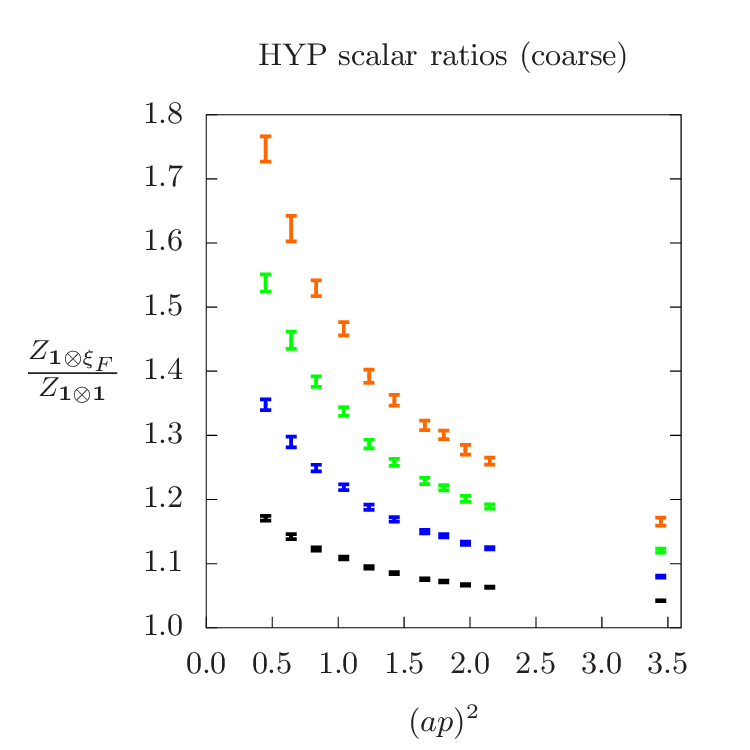}
\caption{As for Fig.~\ref{fig:HYP_coarse_ratios_vectors} but
for the scalar ratios.}
\label{fig:HYP_coarse_ratios_scalars}
\end{center}
\end{figure}

We next display the NPR renormalization-scale dependence of the ratios.
Figures~\ref{fig:HYP_coarse_ratios_vectors}, 
\ref{fig:HYP_coarse_ratios_tensors} and
\ref{fig:HYP_coarse_ratios_scalars} 
show this respectively for the S, T and V ratios on the
coarse lattices.
We omit the axial ratios since they
are very similar to the vectors.
We recall that PT predicts to all orders that the ratios should be independent
of $p$, up to discretization effects at large $(ap)^2$ and non-perturbative
effects at small $p^2$.
We might hope that the window in which such effects would be small
runs (for the coarse lattices) from
$|p|\approx 1\;$GeV ($\Rightarrow (ap)^2\approx 0.4$) 
up to $(ap)^2 \sim 2-3$, i.e.\ the entire width of our dataset.
In fact we find moderate scale dependence for vector ratios
(a doubling of the separation from unity from the high end to
the low end of the range), a significantly smaller dependence
for the tensor ratios, but a very strong dependence for the scalars.

A possible interpretation of these results is as follows.
The curvature at small $(ap)^2$ suggests non-perturbative
effects proportional to powers of $1/p^2$. 
These are largest for the scalar ratios,
and for these the lower edge of the NPR window should be moved 
up to $(ap)^2\approx 1.5$ (corresponding to $|p|\approx 2\;$GeV).
The data above this value can be reasonably well fit by a straight line,
consistent with discretization errors.
Extrapolating to $(ap)^2=0$ removes these discretization errors.
For the vector and tensor ratios the lower edge of the
window can be placed at $(ap)^2\approx 1$ (corresponding to
$|p|\approx 1.6\;$GeV).
This is how many NPR results have traditionally been analyzed
(see, e.g., Ref.~\cite{ishi}).

We do not carry out these extrapolations quantitatively,
because there is clearly an uncertainty introduced by the
choice of fitting window, and we are in this work not aiming
to quote results with a full error analysis. 
Nevertheless, what is clear from the figures is that,
after extrapolation, the overall features found at
$(ap)^2=1.66$ and shown in Fig.~\ref{fig:HYP_ratios_both}
would still hold. 
The only change would be that the ratios would be pushed
further away from unity:
by 15-20\% for V, A and T ratios and by
$\approx 50\%$ for the scalar ratios.
Thus for quantitative agreement at $ap=0$ for V, A and
T ratios one needs to use $\alpha\approx 0.28$,
corresponding to $\mu_0\approx 1.3/a\approx 2.2\;$GeV,
while for scalars one needs $\alpha\approx 0.48$,
corresponding to $\mu_0\approx a/2\approx 0.8\;$GeV.
In the former case the scale is reasonable and the
value of $\alpha$ small enough for reasonable convergence,
but for the scalars the convergence 
of PT is suspect.

It is interesting to
ask whether the $(ap)^2$ corrections are of the expected size.
If the ratios are described approximately by $R(ap=0)[1+ x (ap)^2]$, then,
if we take the relevant scale for cut-off effects to be $\pi/a$,
and assume that the (approximate) improvement of the actions leads to 
a reduction by $\sim \alpha$, then we would
expect $|x|\approx \alpha/\pi^2 \sim 0.03$.
In fact, we find, for example, that
$x\approx -0.015$ for 4-link vector ratios, 
$x\approx -0.007$ for 4-link tensors,
and $x\approx -0.06$ for 3-link scalars.
These are of the expected size or somewhat smaller.
For ratios involving smaller numbers of links, which lie
closer to unity, we see that the slopes, $x$, have yet smaller
magnitudes. For example,
the slope of the 2-link tensor ratios are almost zero.
This suggests that there is an additional suppression
arising from a cancellation of discretization effects 
which follows approximately that for the perturbative corrections.

\begin{figure}[tb!]
\begin{center}
\includegraphics[]{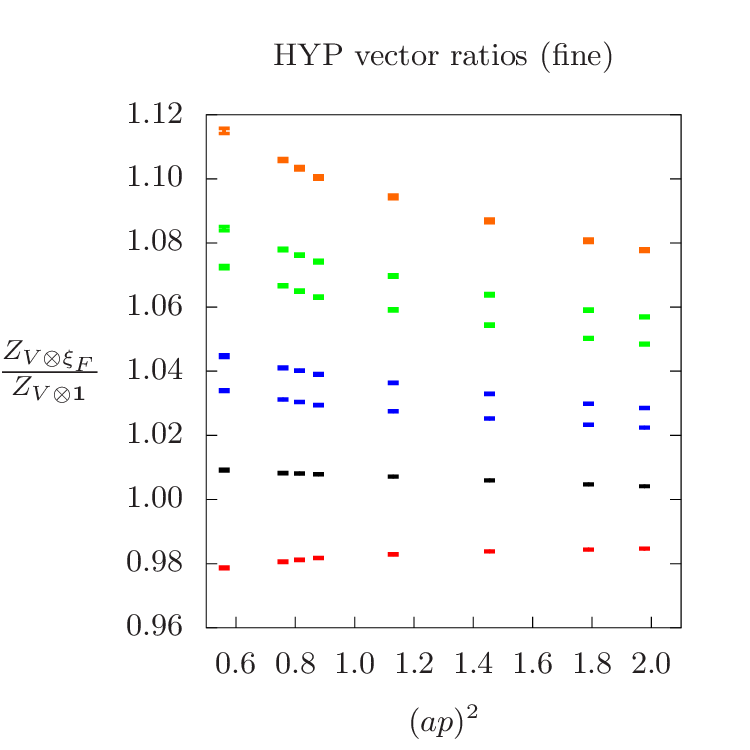}
\caption{Vector ratios vs. $(ap)^2$ for HYP-smeared bilinears
on the fine lattices.
Note that the results at $(ap)^2=0.81$ are
the same as those in Fig.~\protect\ref{fig:HYP_fine_ratios_both}.
Notation as in Fig.~\ref{fig:HYP_coarse_ratios_vectors}.} 
\label{fig:HYP_fine_ratios_vectors}
\end{center}
\end{figure}

The corresponding plots for the fine lattices are qualitatively
similar and, for the sake of brevity, we display only
the results for the vector bilinears 
[Fig.~\ref{fig:HYP_fine_ratios_vectors}].
Note that the range of $(ap)^2$ that is covered is smaller than
on the coarse lattices.
Since the ratio of squared lattice spacings is
$(a_{\text{coarse}}/a_{\text{fine}})^2 \approx 2$, the lower edge of
the NPR window should be halved compared to the coarse lattices
(since it is set by a physical momentum).
Thus the lower edge for V, A and T ratios should move from
$(ap)^2\approx 1$ to $(ap)^2\approx 0.5$,
as a result of which the entire momentum range shown in 
Fig.~\ref{fig:HYP_fine_ratios_vectors} should lie in the window.
This is consistent with our results, which are approximately
linear across the figure. The same is true for the tensor ratios,
while for the scalars the lower edge of the window must be moved up.

Comparing Figs.~\ref{fig:HYP_coarse_ratios_vectors}
and ~\ref{fig:HYP_fine_ratios_vectors}, we see that,
aside from the 1-link (black) points,
all the ratios move towards unity as one goes from
the coarse to the fine lattices at a fixed value of $(ap)^2$.
The 1-link points start very close to unity and remain there.
This same ``collapse towards unity'' occurs for the tensor and
scalar ratios. This is qualitatively what we expect,
because discretization errors should be similar for both
lattice spacings at fixed $(ap)^2$, 
while non-perturbative $1/p^n$ effects should be small 
as we are in the NPR window,
and so the change in the ratios should (if one-loop PT is
reasonably accurate) fall like $\alpha(\mu_0)$ with $\mu_0\sim 1/a$.
In fact, it may be that the discretization errors scale approximately
in this fashion too. 

Pursuing this a little more quantitatively, we find that the values
of the slopes $x$ are approximately the same for corresponding
quantities at the two lattice spacings. 
This holds for all the ratios. The uncertainties in our
estimates are large enough to accommodate a possible factor of
$\alpha(1/a)$ reduction in slope for the fine lattices, but
we do not claim to have found such a reduction.
The approximate equality of slopes implies that the values
after extrapolation to $ap=0$ on the fine lattices remain
closer to unity than the corresponding values on the coarse lattices.
This is what one expects from the perturbative prediction.

\begin{figure}[tb!]
\begin{center}
\includegraphics[]{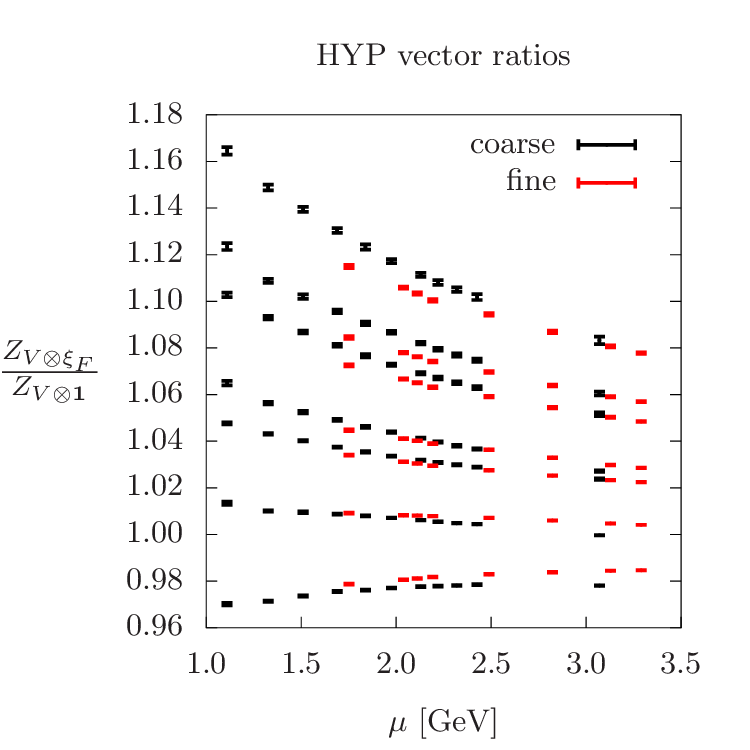}
\caption{Comparison of vector ratios for HYP-smeared bilinears
from coarse (black) and fine (red) lattices, plotted against
$\mu=|p|$.
The coarse results are the same as those presented in 
Fig.~\protect\ref{fig:HYP_coarse_ratios_vectors}.
The link numbers for the coarse results
can be determined by referring to the latter plot; those for
the fine results are the same as for the nearest coarse points.}
\label{fig:HYP_ratios_vectors_redblack}
\end{center}
\end{figure}

\begin{figure}[tb!]
\begin{center}
\includegraphics[]{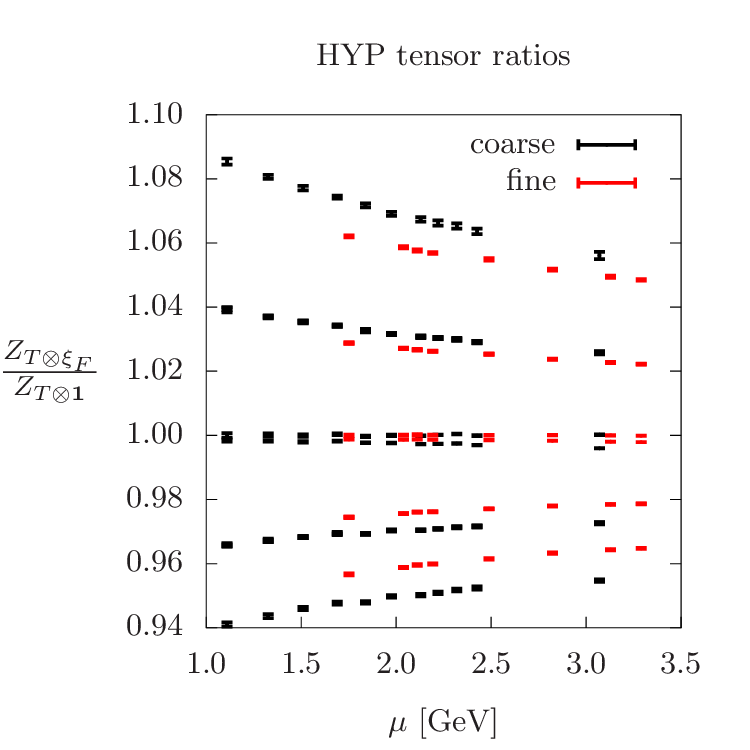}
\caption{As for Fig.~\protect\ref{fig:HYP_ratios_vectors_redblack}
except for tensor ratios.}
\label{fig:HYP_ratios_tensors_redblack}
\end{center}
\end{figure}

\begin{figure}[tb!]
\begin{center}
\includegraphics[]{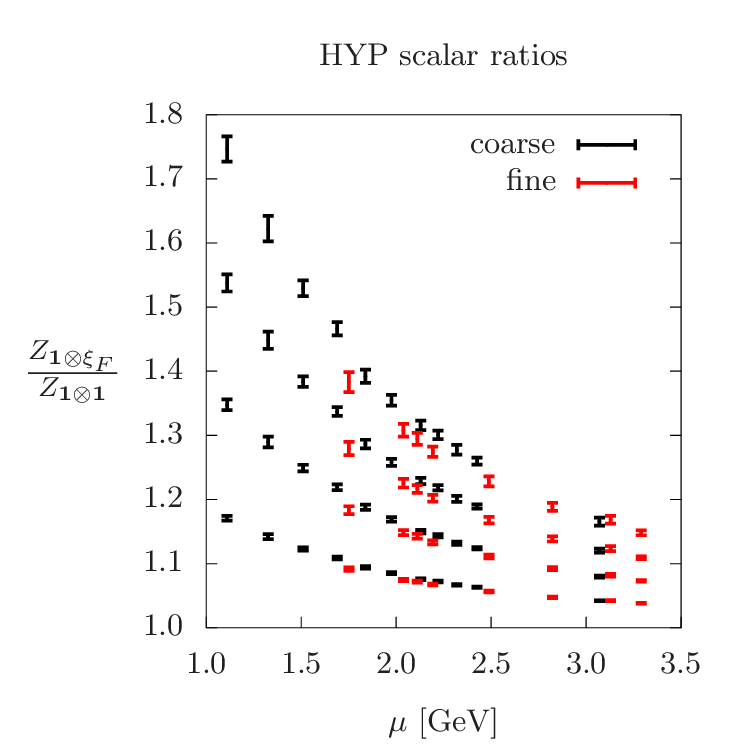}
\caption{As for Fig.~\protect\ref{fig:HYP_ratios_vectors_redblack}
except for scalar ratios.}
\label{fig:HYP_ratios_scalars_redblack}
\end{center}
\end{figure}

Finally, we show,
in Figs.~\ref{fig:HYP_ratios_vectors_redblack},
\ref{fig:HYP_ratios_tensors_redblack} and
~\ref{fig:HYP_ratios_scalars_redblack},
a direct comparison of the results for
the ratios at the two lattice spacings. To make the plots
readable, we plot versus $\mu\equiv |p|$ rather than $(ap)^2$.
This prevents the points from overlapping and distributes them more evenly in the horizontal direction.
The coarse results are identical to those in
Figs.~\ref{fig:HYP_coarse_ratios_vectors},
\ref{fig:HYP_coarse_ratios_tensors}
and \ref{fig:HYP_coarse_ratios_scalars}, respectively,
except that the color coding is no longer used.
A disadvantage of this presentation is that the discretization
errors are, at fixed $\mu$, roughly half as large for the fine
lattice points as for the coarse points. An advantage is that
we expect non-perturbative effects to be similar.
Thus one cannot, from these plots alone, easily disentangle
the perturbative, discretization and non-perturbative contributions.
Nevertheless, one does see the general trend noted above that
the ratios move towards unity on the fine lattices.

\subsection{Ratios for asqtad bilinears}

\begin{figure*}[tb!]
\begin{center}
\includegraphics[]{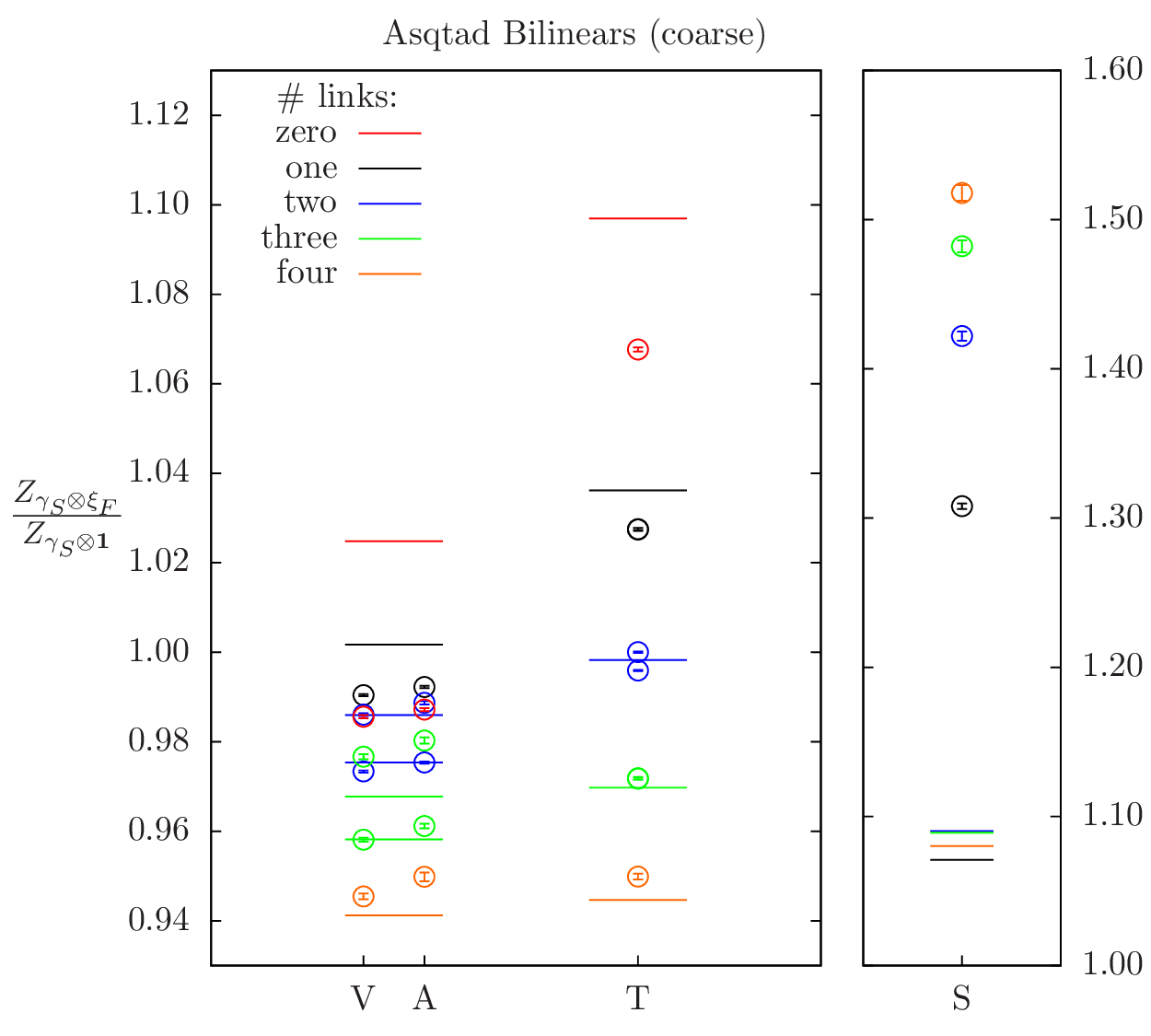}
\caption{As for Fig.~\ref{fig:HYP_ratios_both} but for asqtad fermions,
and with only mean-field improved perturbative predictions shown.
}
\label{fig:ASQ_ratios_both}
\end{center}
\end{figure*}

\begin{figure*}[tb!]
\begin{center}
\includegraphics[]{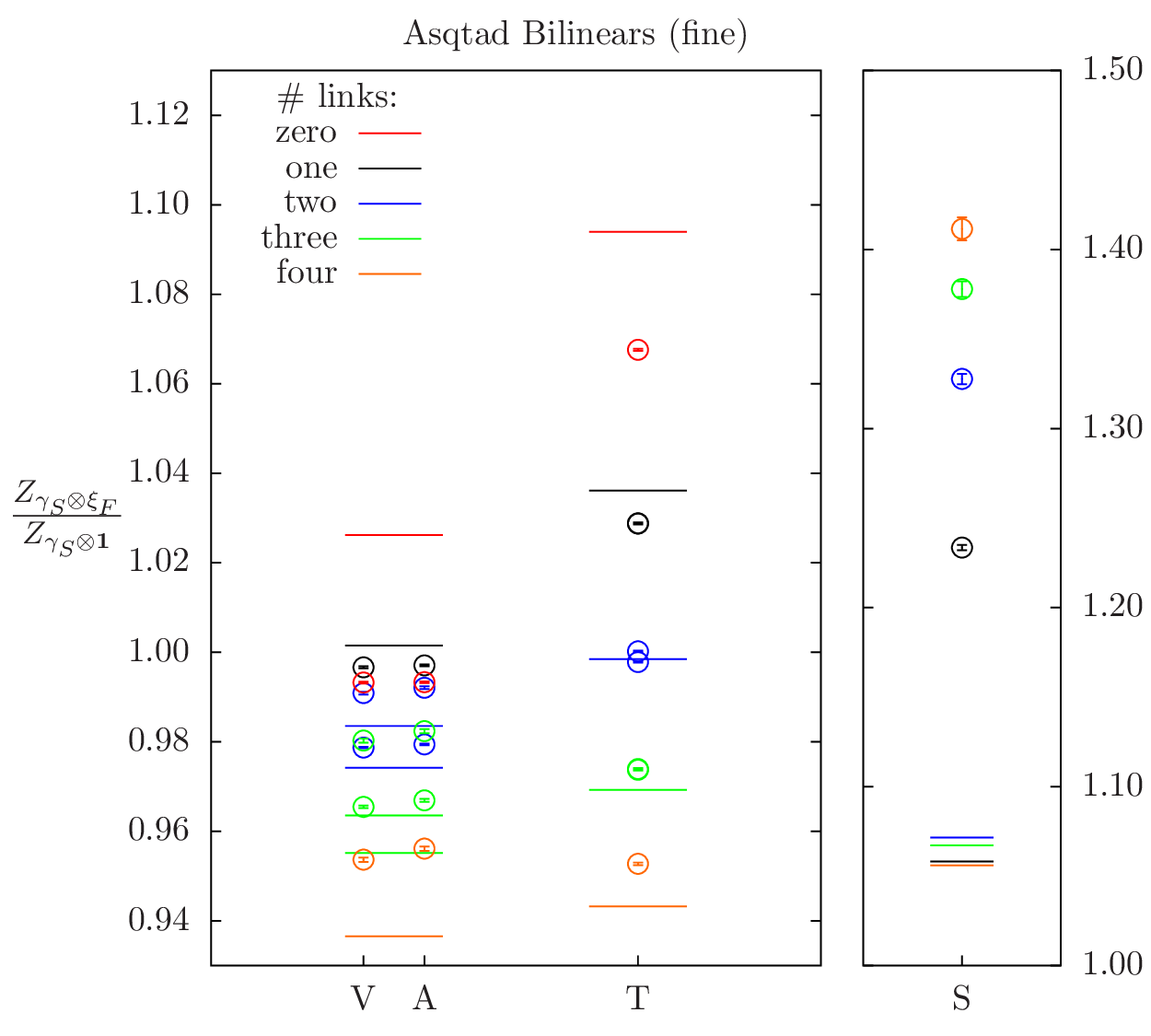}
\caption{As for Fig.~\ref{fig:ASQ_ratios_both},
but for the fine lattices.
}
\label{fig:ASQ_fine_ratios_both}
\end{center}
\end{figure*}

We now turn to the asqtad bilinears, for which the results
turn out to be less well represented by PT,
and harder to understand.
We begin with plots of all ratios at our canonical momentum
on the coarse and fine lattices, Figs.~\ref{fig:ASQ_ratios_both}
and \ref{fig:ASQ_fine_ratios_both}.
Mean-field
improvement is necessary to obtain even reasonably accurate predictions,
so we show only the corresponding results.
The greater importance of mean-field improvement
for asqtad fermions is related to the result
that the corresponding fat links (``Fat7 + Lepage'') have 
traces that are significantly further from unity than the HYP-smeared
links indicating larger fluctuations.
For example, on the coarse lattices,
$\tilde u_0^{\rm ASQ}=1.053 $ to be compared to $\tilde u_0^{\rm HYP}=0.984$.

The asqtad results differ in several noteworthy ways from those
with HYP-smeared fermions.
First, the ordering of the tensor bilinears by link number is
reversed. This is also true for the 2-, 3- and 4-link V and A spins. 
This can be qualitatively understood as follows.
The average of the smeared link in the asqtad operators is larger than
unity.\footnote{%
This is possible because the Fat7 $+$ Lepage links are linear combinations of different paths.}
Bilinear matrix elements 
are thus expected to grow with the number of links, leading to
$Z$-factors which must decrease to compensate.
This is the same argument used above for the HYP-smeared bilinears,
except in that case it leads to the opposite ordering because 
$\tilde u_0^{\rm HYP}<1$.
Here, however, the argument fails for the scalars, which 
have the same ordering as for HYP fermions 
(although they are, in relative terms, more bunched together). 
Of course this argument is naive, as there are correlations between
fluctuations in the links, something that is approximately accounted
for by PT. Indeed, mean-field improved PT does predict the
observed ordering for tensor ratios and the 2-4 link V and A spins. 
Nevertheless, the gross structure
is reproduced in its entirety only for the tensor ratios,
with the predictions for the scalars simply being poor.
This situation is not improved 
by changes to the value of $\alpha$.

On the positive side we note that the all-orders predictions of
degeneracy are borne out at about the same level as for HYP-smeared
operators. In addition, the quantitative disagreements with PT for the
V, A and T ratios are at the few percent level, which could be 
understood as generic ${\cal O}(\alpha^2)$ effects.

Finally, we comment on the changes as one goes from coarse to fine
lattices, which exhibit a more complicated pattern than for the HYP-smeared
operators. The NPR results for the scalar, vector and axial ratios 
do move towards unity as $a$ decreases (as in the HYP case), 
but the tensor ratios are almost unchanged.
The perturbative predictions for scalar ratios move toward unity,
while those for vector and axial ratios are almost unchanged, and
the predictions for tensors move slightly {\em away} from unity.
This complicated pattern of perturbative predictions is due to
the use of mean-field improvement and the fact that the non-perturbative
value of $\tilde u_0^{\rm ASQ}$ drops by a smaller factor from coarse
to fine lattices than is predicted by perturbation theory.
For example,
evaluating $\alpha$ at the scale $1/a$, 
Eq.~(\ref{eq:u0ASQinPT}) predicts 1.12 and 1.10 for coarse and fine
lattices, respectively, to be compared to the measured values 1.053 and
1.051.

\begin{figure}[tb!]
\begin{center}
\includegraphics[]{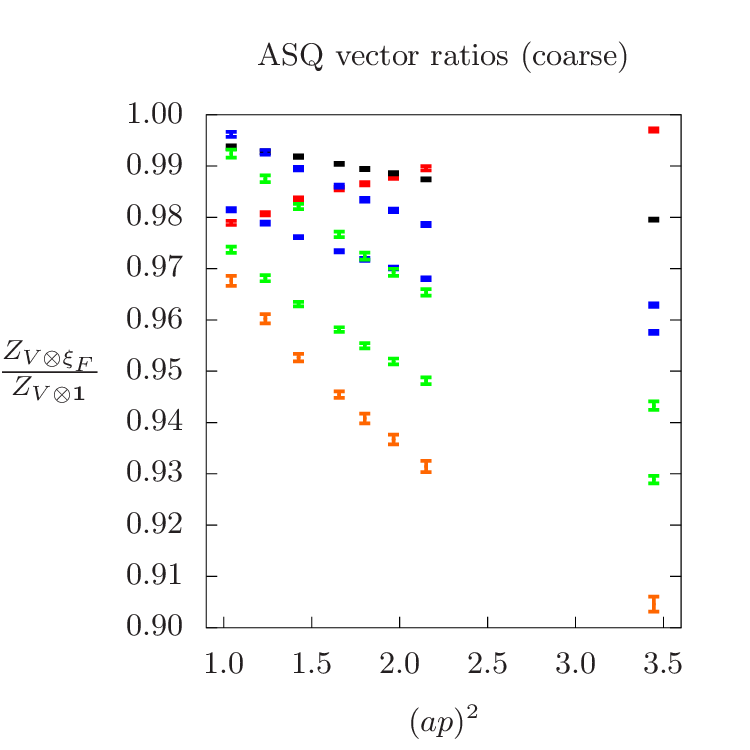}
\caption{Scale dependence of asqtad vector ratios on coarse MILC lattices.
Notation as in Fig.~\ref{fig:HYP_coarse_ratios_vectors}.}
\label{fig:ASQ_coarse_ratios_vectors}
\end{center}
\end{figure}

\begin{figure}[tb!]
\begin{center}
\includegraphics[]{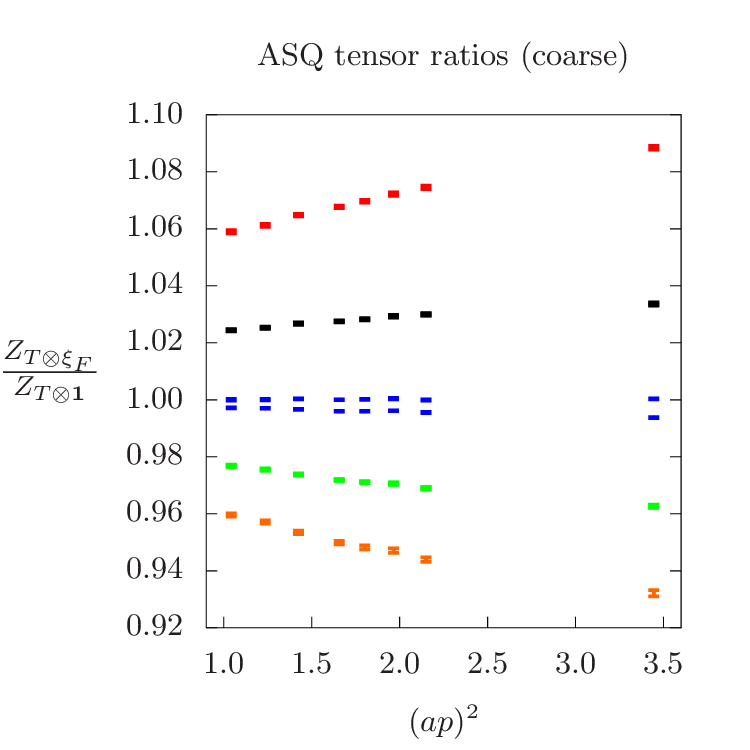}
\caption{As for Fig.~\ref{fig:ASQ_coarse_ratios_vectors} but
for tensor ratios.}
\label{fig:ASQ_coarse_ratios_tensors}
\end{center}
\end{figure}

\begin{figure}[tb!]
\begin{center}
\includegraphics[]{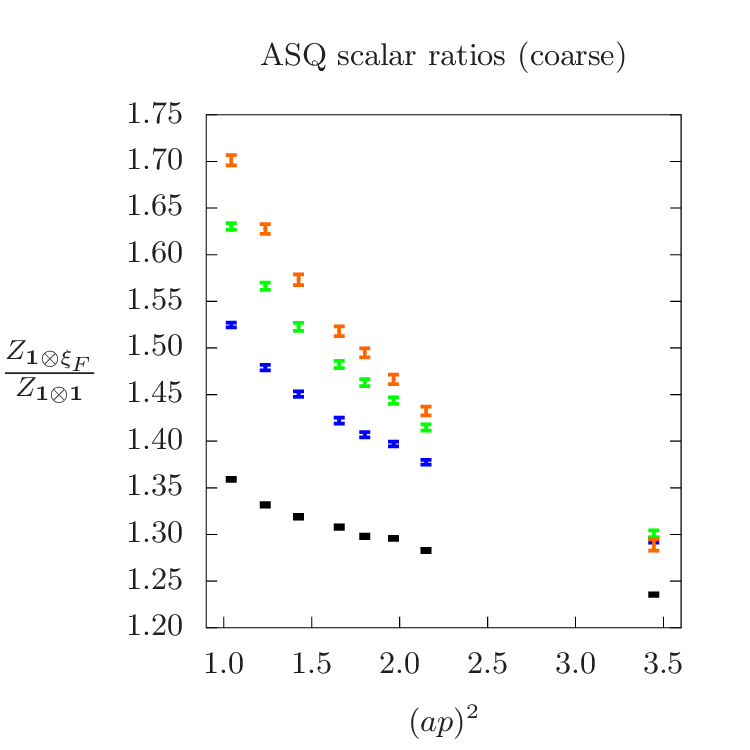}
\caption{As for Fig.~\ref{fig:ASQ_coarse_ratios_vectors} but
for scalar ratios.}
\label{fig:ASQ_coarse_ratios_scalars}
\end{center}
\end{figure}

\begin{figure}[tb!]
\begin{center}
\includegraphics[]{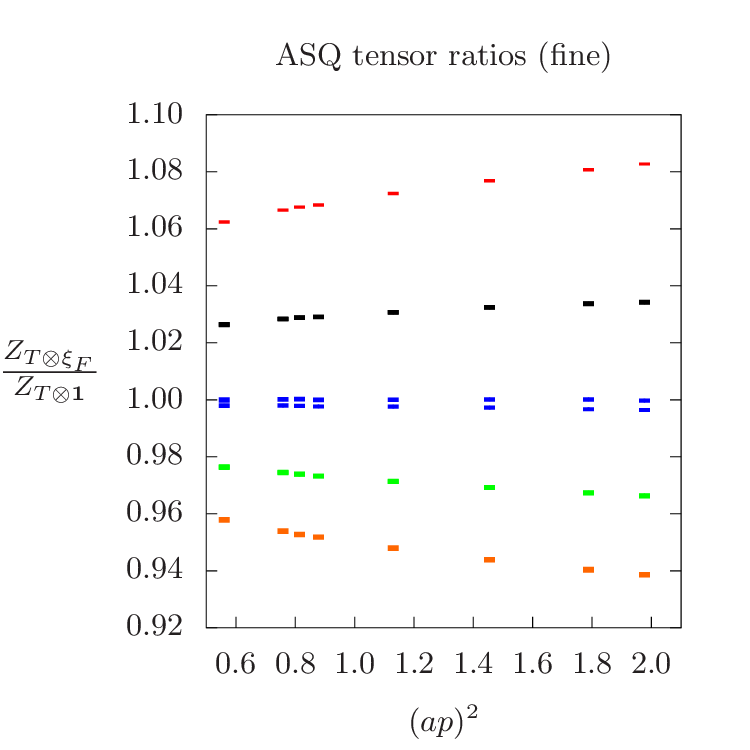}
\caption{Tensor ratios vs. $(ap)^2$ for HYP-smeared bilinears
on the fine lattices.
Note that the results at $(ap)^2=0.81$ are
the same as those in Fig.~\protect\ref{fig:ASQ_fine_ratios_both}.
Notation as in Fig.~\ref{fig:ASQ_coarse_ratios_tensors}.} 
\label{fig:ASQ_fine_ratios_tensors}
\end{center}
\end{figure}

The momentum dependence of the asqtad ratios on the coarse
lattices are shown in Figs.~\ref{fig:ASQ_coarse_ratios_vectors},
\ref{fig:ASQ_coarse_ratios_tensors}
and \ref{fig:ASQ_coarse_ratios_scalars}, with
one example (the tensors) of the corresponding behavior on the fine lattices
shown in Fig.~\ref{fig:ASQ_fine_ratios_tensors}.
Note that on the coarse lattices
the range of $(ap)^2$ is smaller than in the
corresponding HYP plots, so any curvature due to
non-perturbative effects will be harder to see.
This explains why the curves for V and T spins appear
more linear. For these cases the NPR window appears to
cover the entire range of our data
(consistent with the results from HYP-smeared bilinears), 
while for the scalars the lower cut-off again needs to be moved up
to $(ap)^2\sim 1.5$ on the coarse lattices.

We first discuss the vector and tensor ratios.
Although the plots look superficially
different from those with HYP-smeared fermions,
we note that all the slopes have the same signs in the two cases
(comparing data with the same number of links), 
and indeed have the same ordering of magnitudes. 
The only change in the discretization effects between 
HYP and asqtad cases is that the slope-coefficients $x$ are
about twice as large for asqtad bilinears.
This is consistent with the general experience that HYP 
smearing leads to smaller discretization effects. 
We also find that, as for HYP-smeared bilinears, 
the slopes at the two lattice spacings are similar.

For tensor ratios, the ordering seen in
Fig.~\ref{fig:ASQ_ratios_both} remains valid over our entire
momentum range, and also after extrapolation to $ap=0$.
Thus by a small rescaling of $\alpha$ one can retain
quantitative agreement with one-loop PT at $ap=0$.
On the other hand, extrapolating the fine lattice
results to $ap=0$ leads to values which lie a little
{\em further} from unity 
[cf.\ Figs.~\ref{fig:ASQ_coarse_ratios_tensors}
and  \ref{fig:ASQ_fine_ratios_tensors}],
which is not consistent with PT.

For the vector ratios extrapolation to $ap=0$
reshuffles the ordering, with the zero-link ratio
now having the smallest $Z$-factor.
Thus the perturbative predictions of Fig.~\ref{fig:ASQ_ratios_both}
become worse for the vectors after extrapolation,
even after possible rescalings of $\alpha$.

The $ap$ dependence of the scalar ratios, by contrast,
is similar to that for the HYP-smeared bilinears.
The ordering is maintained by extrapolations to $ap=0$,
with slope-coefficients that are similar 
(not differing by a factor of two).
However, the already very poor perturbative predictions 
become even worse after the extrapolations.

A perplexing feature of the results for momentum dependence
is that, at the highest values of $(ap)^2$,
both vector and scalar ratios become much closer to the
perturbative predictions, particularly in terms of the ordering
and relative splittings. We do not understand why this should be.

For the sake of brevity, we do not show the direct comparisons
of coarse and fine asqtad ratios versus $|p|$. These plots
are both messy and hard to interpret, adding little to the
preceding discussion.

In summary, one-loop PT fails to provide
 even a qualitative description of many of the
features observed for the asqtad bilinears,
with the exception of the tensor ratios. 
One should keep in mind, however, that the
disagreements with the vectors are well within the expected size
of generic two-loop contributions.

\subsection{Results for denominators}

Finally we turn to a discussion of the denominators in the
ratios, namely the matching factors $Z_{1\otimes1}$, $Z_{\gamma_\mu\otimes1}$ and
$Z_{\gamma_\mu\gamma_\nu\otimes1}$ 
which we label simply
$Z_S$, $Z_V$ and $Z_T$, respectively.
Unlike the ratios, these
quantities have anomalous dimensions (even $Z_V$ in the RI$'$ scheme),
so that they do depend on $\mu=|p|$
even in the absence of non-perturbative effects and discretization errors.
This dependence is described in perturbation theory by
the result Eq.~(\ref{eq:Zmaster}), the ingredients for which
are collected in appendices~\ref{app:pert} and \ref{app:running}.
In brief, one runs in the continuum (in the RI$'$ scheme) from
$\mu$ to $\mu_0\sim 1/a$, and then matches to the lattice scheme
at that scale. This matching is done, for technical reasons,
using the $\overline{\text{MS}}$ scheme as an intermediate step.

\begin{figure}[h]
\begin{center}
\includegraphics[]{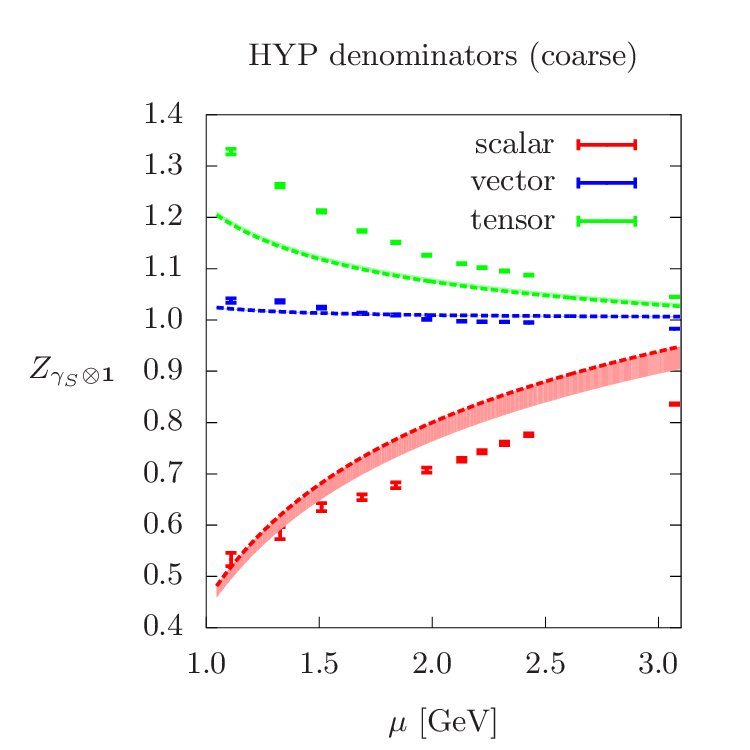}
\caption{Comparison of scale dependence
of the HYP-smeared taste-singlet scalar, vector, 
and tensor $Z$-factors computed non-perturbatively on the coarse lattices
to the perturbative prediction described in the text.
The colored bands give the variation in the perturbative
prediction arising from varying the intermediate matching scale
between between $1/a$ (dotted line) and $2/a$.}
\label{fig:HYP_coarse_denominators}
\end{center}
\end{figure}

\begin{figure}[h]
\begin{center}
\includegraphics[]{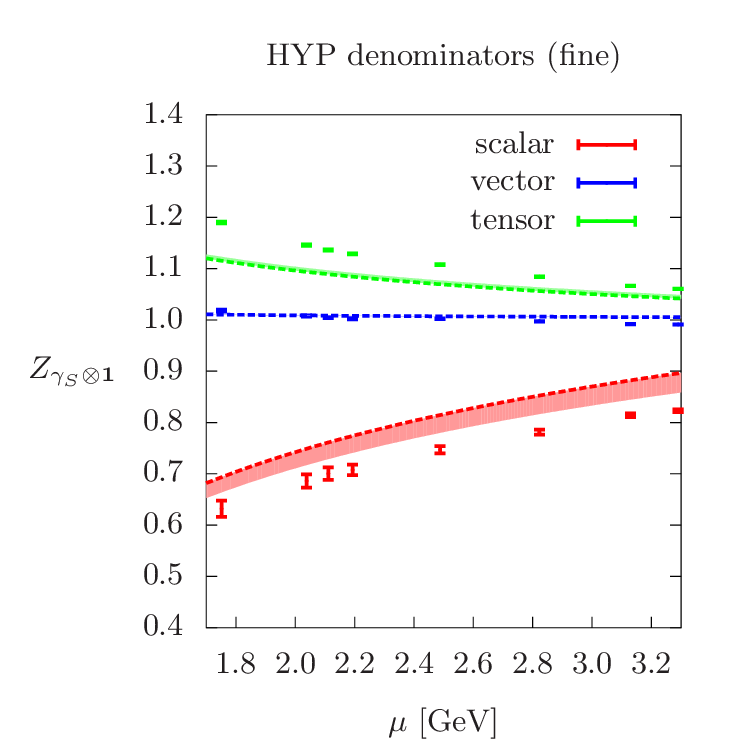}
\caption{As for Fig.~\ref{fig:HYP_coarse_denominators} but
on the fine lattices.}
\label{fig:HYP_fine_denominators}
\end{center}
\end{figure}

As above, we consider first the HYP-smeared bilinears.
Results from the coarse and fine lattices are shown in
Figs.~\ref{fig:HYP_coarse_denominators}
and ~\ref{fig:HYP_fine_denominators}, respectively.  
Note again that the range of $\mu$ differs in the two cases.
For the perturbative results, we use the non-mean-field improved
result (which lies very close to the mean-field improved result),
and display a band to give
an indication of the uncertainty due to truncation errors.
This is obtained by varying $\mu_0$ between $1/a$ and $2/a$,
a range for which $\Delta \alpha \approx \alpha^2$.
We stress that the  weakest link in the perturbative result
is the one-loop matching between
the lattice and $\overline{\text{MS}}$ schemes;
all other running or matching is done at 3 or 4-loop order.
We also note that, as for the $Z$-factor ratios, this
estimate of truncation errors is not the most conservative
when $Z$ is close to unity, 
because there can be generic ${\cal O}(\alpha^2)$
terms of size 5-9\%.

The figures show good qualitative
agreement between the NPR and PT results in all three channels. 
The ordering is correct and the $\mu$ dependence is reasonably
well predicted. Quantitatively the perturbative
prediction undershoots the separation from unity for $Z_T$ and
$Z_V$, even allowing for the predicted uncertainty band. 
This mismatch is small enough, however, that it could be due to
generic two-loop contributions.
The level of quantitative agreement is somewhat worse than that
found above for the ratios: 
for these, PT could reproduce all the vector and tensor ratios
with choices of $\mu_0$ lying in the range $1/a-2/a$.

Unlike the ratios,
the $Z$-factors do not themselves have a good continuum limit,
due to the non-vanishing anomalous dimensions.
To take a continuum limit one must multiply them by hadronic matrix
elements of the corresponding bilinears, which we do not have
available here. Because of this, there is no general
expectation that results from the fine lattices should 
lie closer to unity than those from the coarse lattices,
even ignoring discretization errors.
What one might expect, however, is that
the perturbative prediction should become more accurate, since
the intermediate matching scale $\mu_0$ is higher.
We do in fact see a small improvement
between Fig.~\ref{fig:HYP_coarse_denominators} and 
\ref{fig:HYP_fine_denominators}.

\begin{figure}[tbp!]
\begin{center}
\includegraphics[]{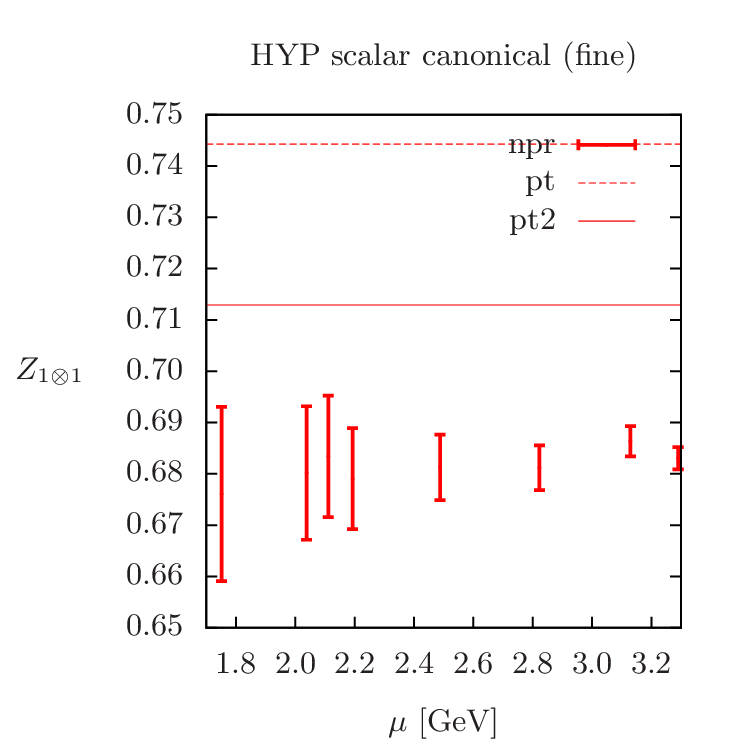}
\caption{Predictions for $Z_S^{\text{RI'}}(2 \; \text{GeV})$ 
with HYP-smeared fermions on fine MILC lattices.
The $Z$-factor is run from NPR scale $\mu$ 
to 2 GeV using continuum perturbation theory.
The perturbative predictions use an intermediate conversion scale
of $1/a$ (``pt'') or $2/a$ (``pt2'').}
\label{fig:HYP_fine_denominators_scalar_canonical}
\end{center}
\end{figure}

\begin{figure}[tbp!]
\begin{center}
\includegraphics[]{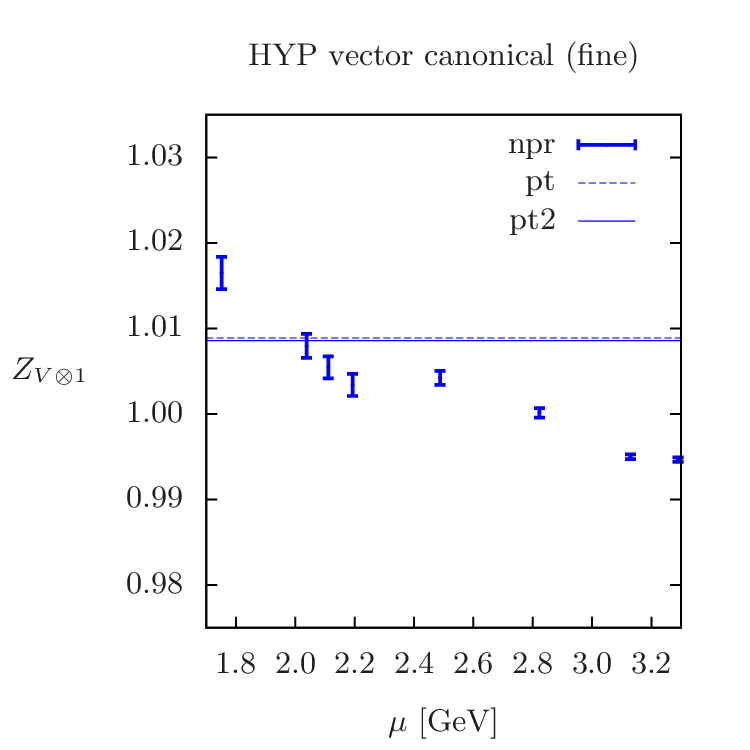}
\caption{As for
Fig.~\ref{fig:HYP_fine_denominators_scalar_canonical}
but for $Z_V$.
}
\label{fig:HYP_fine_denominators_vector_canonical}
\end{center}
\end{figure}

\begin{figure}[tbp!]
\begin{center}
\includegraphics[]{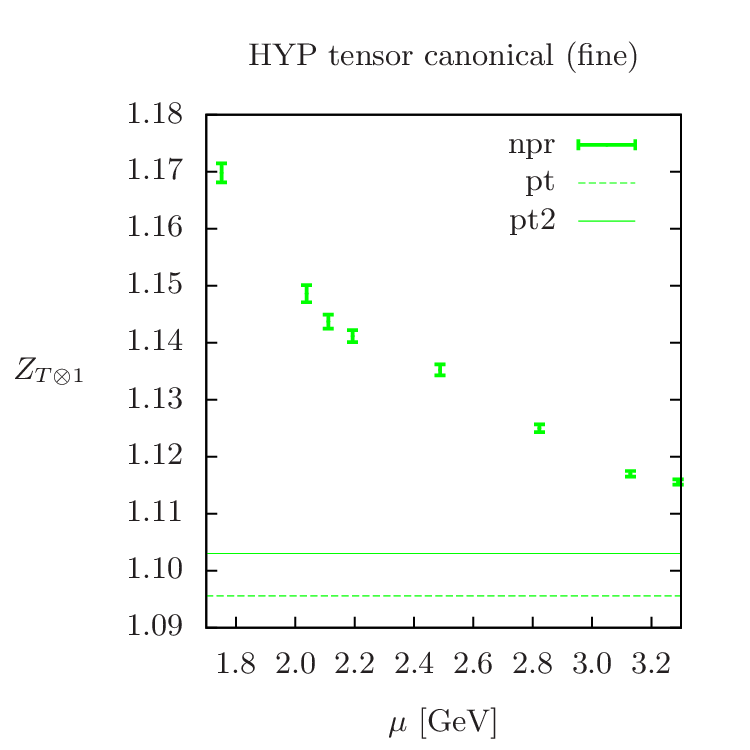}
\caption{As for
Fig.~\ref{fig:HYP_fine_denominators_scalar_canonical}
but for $Z_T$.}
\label{fig:HYP_fine_denominators_tensor_canonical}
\end{center}
\end{figure}

In order to disentangle the predicted running with $\mu$ from
discretization effects, we can run our results in the RI$'$ scheme
from $\mu$ to a canonical scale which we choose to be $2\;$GeV.
This running is done at three or four loop order using
continuum anomalous dimensions (see app.~\ref{app:running}).
The hope is that the data will significantly ``flatten'',
leaving a residual $(a\mu)^2$ dependence.
In Figs.~\ref{fig:HYP_fine_denominators_scalar_canonical},
\ref{fig:HYP_fine_denominators_vector_canonical} and
\ref{fig:HYP_fine_denominators_tensor_canonical}
we show results after this running for the fine lattices.
Results on the coarse lattices are similar and are not shown.
The perturbative predictions are obtained as above, but with
$p$ replaced by $2\;$GeV.
We recall that the NPR window covers the range of results
shown in these plots.

For $Z_S$, we see that the flattening is successful, although
the perturbative prediction for the absolute value misses the data.
Nevertheless, a generic two-loop term would be sufficient to make up
the gap.
$Z_V$ varies by $\approx 1-2\%$ over the momentum range shown,
and, if extrapolated to $a\mu=0$, will lie quite close to the
perturbative prediction.
$Z_T$ varies more rapidly, and, if extrapolated linearly in
$(ap)^2$ to $a=0$, will become $\approx 1.16$.
This is $\sim 5\%$ above the perturbative prediction, a difference
which could be bridged by two-loop perturbative contributions.
The slope-coefficient is $x\approx -0.023$,
and is comparable to that for ratios.

\begin{figure}[tb!]
\begin{center}
\includegraphics[]{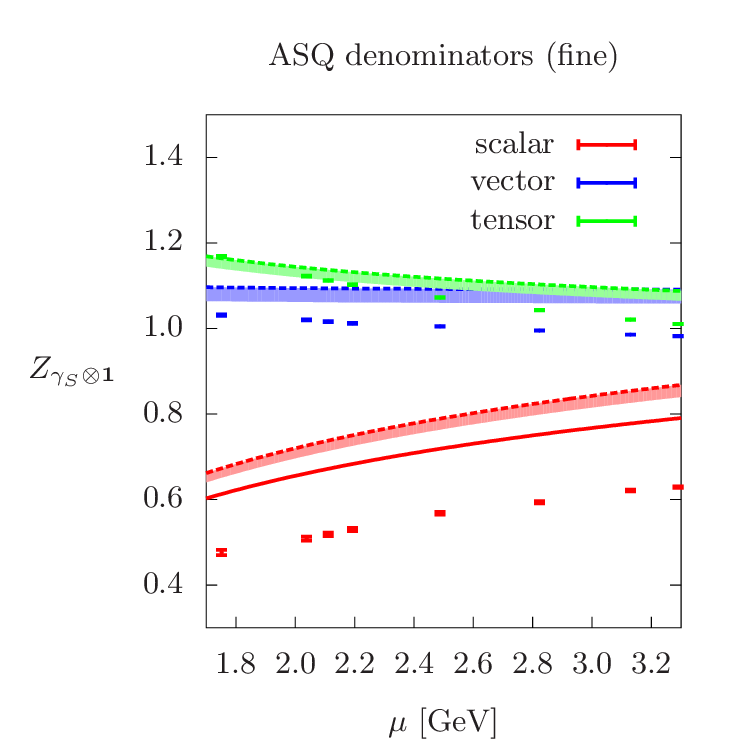}
\caption{Comparison of the asqtad $Z_S$, $Z_V$ and
$Z_T$ on fine lattices with mean-field improved PT.
Details as in Fig.~\ref{fig:HYP_fine_denominators},
except that the two-loop prediction for $Z_S$ (red solid line)
is also shown.}
\label{fig:ASQ_fine_denominators}
\end{center}
\end{figure}

We now turn to the asqtad denominators, for which we show
the running with NPR scale on the fine lattices in
Fig.~\ref{fig:ASQ_fine_denominators}. 
Results are similar on the coarse lattices.
We compare here to mean-field improved perturbation theory,
since without mean-field improvement the asqtad ratios 
are poorly represented, as discussed above. 
We note that mean-field improvement impacts
the predictions for $Z_V$ and $Z_T$, but not that for $Z_S$.
For $Z_S$, we also show the perturbative result including
the two-loop lattice to $\overline{\rm MS}$ matching factor
from Ref.~\cite{MILC2loop}. This has a much weaker dependence
on the intermediate matching scale than that using
one-loop matching,
and we show the result only for intermediate scale $1/a$.
We stress that one cannot directly gauge the rate of convergence
of the perturbative series from a comparison of ``one-loop''
and ``two-loop'' results, since both are composed of
several components, some of which are being evaluated at three 
or four loop order
[see Eq.~(\ref{eq:Zmaster})]. What one can see, however, is that
shift between ``one-loop'' and ``two-loop'' results is of the
$\sim 5\%$ size expected of a generic two-loop term on the fine lattices.
Compared to the corresponding HYP-smeared results
(Fig.~\ref{fig:HYP_fine_denominators}), we observe that
the NPR result for $Z_S$ is much further from unity,
and also further from the perturbative predictions.

\begin{figure}[tb!]
\begin{center}
\includegraphics[]{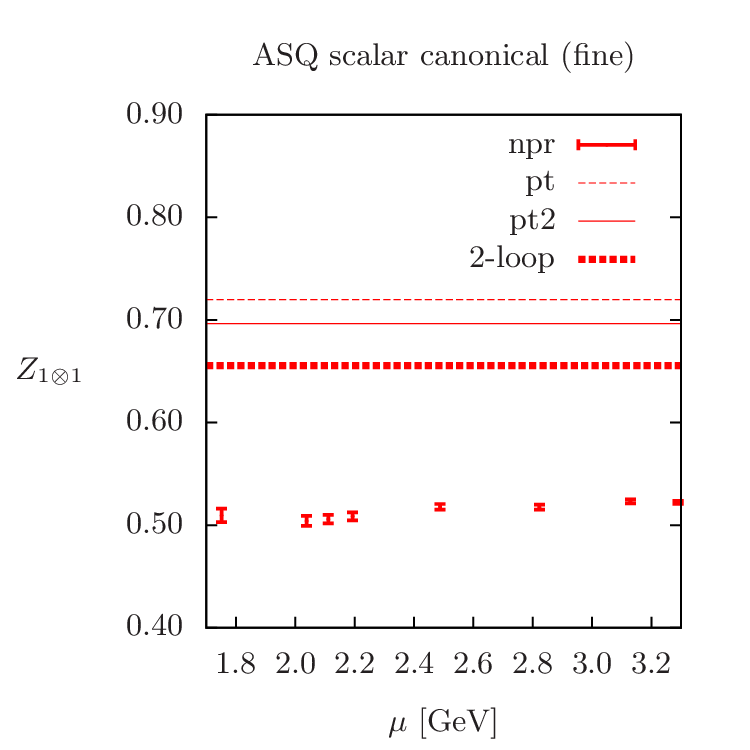}
\caption{Predictions for $Z_S^{\text{RI'}}(2 \; \text{GeV})$ 
with asqtad fermions on fine MILC lattices.
The $Z$-factor is run from NPR scale $\mu$ 
to 2 GeV using continuum perturbation theory.
The one-loop perturbative predictions use an intermediate conversion scale
of $1/a$ (``pt'') or $2/a$ (``pt2''). 
Also shown is the two-loop
perturbative prediction (with intermediate scale $1/a$).}
\label{fig:ASQ_fine_denominators_scalar_canonical}
\end{center}
\end{figure}

\begin{figure}[tb!]
\begin{center}
\includegraphics[]{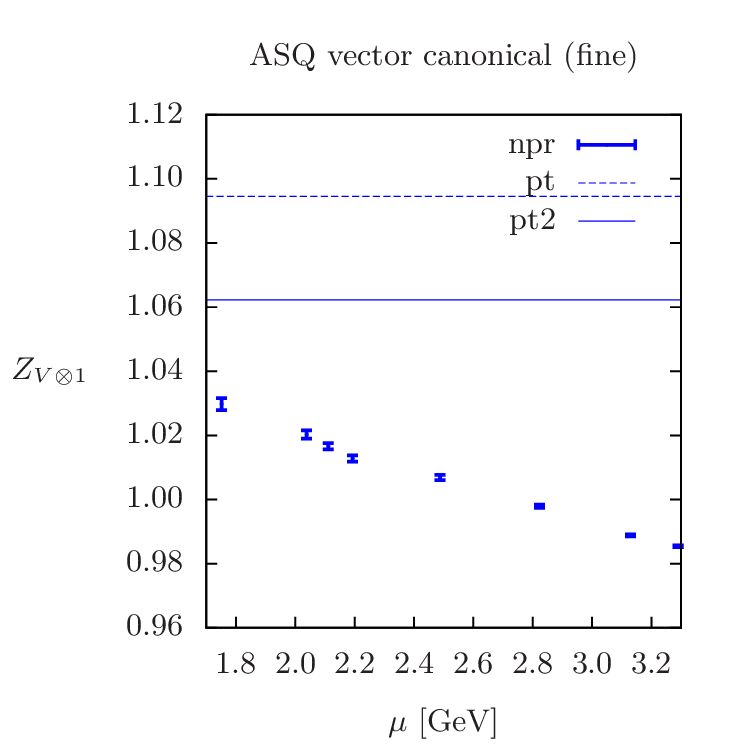}
\caption{As for Fig.~\ref{fig:ASQ_fine_denominators_scalar_canonical}
except for $Z_V$.}
\label{fig:ASQ_fine_denominators_vector_canonical}
\end{center}
\end{figure}

\begin{figure}[tb!]
\begin{center}
\includegraphics[]{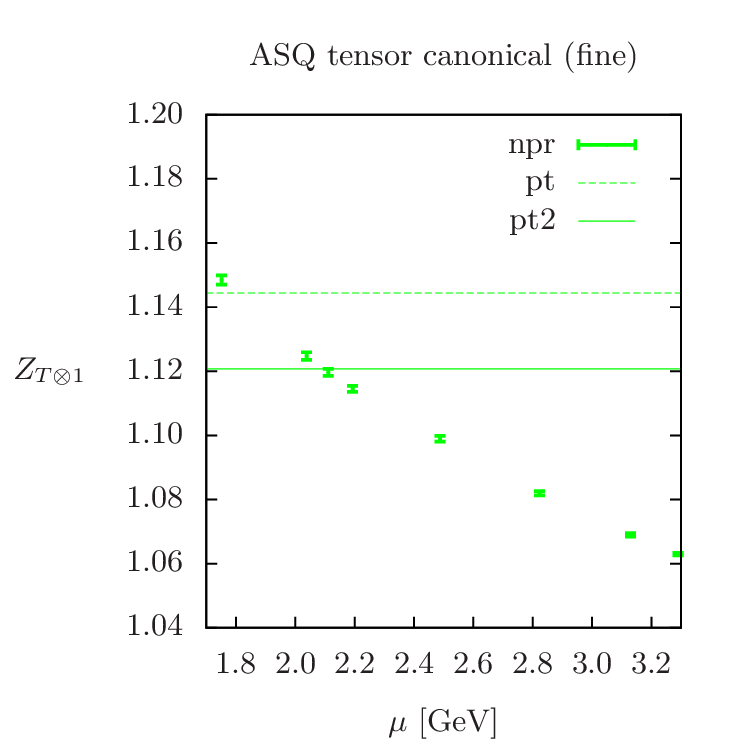}
\caption{As for Fig.~\ref{fig:ASQ_fine_denominators_scalar_canonical}
except for $Z_T$.}
\label{fig:ASQ_fine_denominators_tensor_canonical}
\end{center}
\end{figure}

We learn more from the results after flattening, shown in 
Figs.~\ref{fig:ASQ_fine_denominators_scalar_canonical},
~\ref{fig:ASQ_fine_denominators_vector_canonical}
and~\ref{fig:ASQ_fine_denominators_tensor_canonical}.
For $Z_S$, it is striking that (as for the HYP bilinears),
the results show little indication of $(a\mu)^2$ effects,
indicating that the four-loop anomalous dimension is
giving a good representation of the $\mu$ dependence.
On the other hand, the value itself lies $\sim 0.2$ below
the ``one-loop'' and $\sim 0.15$ below the ``two-loop''
predictions, indicating a failure of convergence since this
gap is too large to be bridged by a generic 
${\cal O}(\alpha^3)$ term.
We note that this gap is the reason why, as described in
the Introduction, the value of 
$m_s$ obtained from our NPR results lies significantly
above that obtained using two-loop matching.
Specifically, on the fine lattices, NPR yields
$m_s=105\;$MeV~\cite{Lytlethesis}
while the two-loop result is $86\;$MeV~\cite{MILC2loop}.

The situation is much better for $Z_V$ and $Z_T$.
For $Z_V$ there is a mild $\mu$ dependence, which brings
the result at $a\mu=0$ close to the one-loop prediction.
For $Z_T$, the $\mu$ dependence is somewhat stronger and
leads to a value at $a\mu=0$ of $Z_T(a\mu=0)\approx 1.18$,
within 5\% of the perturbative prediction.
In both cases, the gap can be bridged by a generic two-loop
contribution.

\section{Conclusions}
\label{sec:conc}

We have implemented non-perturbative renormalization for
general staggered-fermion bilinear operators, using a method
that is consistent with the symmetries of the staggered action.
We have shown how those symmetries constrain the 
propagator and vertex functions to have the expected
continuum forms at leading non-trivial
order in an expansion in the lattice spacing.
We have also introduced ``covariant bilinears'', which
transform irreducibly under the lattice symmetries and thus
do not mix, unlike the traditional ``hypercube bilinears''.

We have calculated $Z$-factors for 30 different operators
having spins V, A, T and S. It is well known that, for unimproved
staggered fermions, many $Z$-factors, particularly those for scalars,
lie very far from unity and have perturbative expansions which
are not convergent~\cite{ishi,PS}. We have rechecked this result
ourselves. It is also well known that these problems can be
substantially improved using smeared lattice links and other
forms of action improvement. Here we have used HYP-smeared 
and asqtad fermions. By studying many operators we are able
to give a general judgement on the utility of perturbation theory
for these two types of fermion. A useful tool in this regard is
the use of ratios for which the overall running due to 
anomalous dimensions cancels, allowing a study of the 
approach to the continuum limit.

Overall, we find that the HYP-smeared $Z$-factors lie relatively
close to unity and can be predicted by one-loop PT as long
as one includes a generic uncertainty of relative size
${\cal O}(1) \times \alpha(1/a)^2$. This holds both for ratios and
for the $Z$-factors themselves. 
In fact, PT works more accurately than this
for the vector and tensor ratios,
with an uncertainty given by the square of the one-loop term sufficing.
The detailed ordering of these ratios is predicted very well.
We also find that discretization errors proportional
to $(ap)^2$ are of the expected size or smaller.

For the asqtad bilinears, one-loop PT is less successful.
Only for the tensor ratios does it approach the efficacy observed
in the HYP-smeared case, while for the scalars there appears to
be a breakdown in convergence. 

These results have implications for extracting physical predictions
from staggered simulations. The recent calculation of $B_K$
using HYP-smeared fermions used one-loop perturbative results for
the needed $Z$-factors~\cite{SW-1,SW11}. 
The anomalous dimension of the operator
which appears is roughly comparable to that for the tensor bilinear,
and thus we can use the latter as a guide to how well one-loop PT
reproduces the $Z$-factor obtained using NPR. We find in this case
(see, e.g., Figs.~\ref{fig:HYP_coarse_denominators}
and \ref{fig:HYP_fine_denominators})
that one-loop PT gives a good estimate as long as one uses
an error estimate of ${\cal O}(1)\times \alpha^2$.
This is, in fact, the estimate used in Ref.~\cite{SW-1,SW11}. 

As for quark masses obtained using one- or two-loop perturbative
matching, the results of Figs.~\ref{fig:ASQ_fine_denominators}
and \ref{fig:ASQ_fine_denominators_scalar_canonical} show
that there is a substantial gap between the perturbative and
non-perturbative results for $Z_S=1/Z_m$ with asqtad fermions.
This gap is larger than a straightforward estimate of the truncation error.
This suggests that the systematic error in the quark masses
obtained in Refs.~\cite{MILC1loop,MILC2loop,milcrmp} 
may be larger than previously estimated.
To study this point further, it will be important to use NPR with
non-exceptional momenta~\cite{andrewlat12}.

Finally, we note that present large-scale simulations with
staggered fermions now use HISQ rather than asqtad quarks.
HISQ quarks combine the advantages of HYP smearing with the
full ${\cal O}(a^2)$ improvement of asqtad quarks (and in addition
reduce discretization errors for heavier quarks)~\cite{hisq}.
Thus we expect the success of PT in describing $Z$-factors
for HYP-smeared operators to carry over to operators composed
of HISQ quarks.

\section*{Acknowledgments}

Computations for this work were carried out on USQCD Collaboration
clusters at Fermilab. 
The USQCD Collaboration is funded by the Office of Science of the U.S. 
Department of Energy.
The work of AL was supported in part by STFC grants ST/G000557/1 and ST/J000396/1.
The work of SS is supported in part by the U.S. 
Department of Energy grant no.~DE-FG02-96ER40956. 

%
\appendix

\section{Notation and conventions}
\label{app:notation}

\subsection{Staggered matrix conventions}
We use the notation of Refs.~\cite{DK,DS}, which introduce
two sets of matrices unitarily equivalent to the general spin-taste
matrices. A basis for the latter is $(\gamma_S\otimes \xi_F)$, 
where a general spin matrix is labeled by the hypercube vector $S$,
\begin{equation}
\gamma_S=\gamma_1^{S_1}\gamma_2^{S_2}\gamma_3^{S_3}\gamma_4^{S_4},
\end{equation}
while the general taste matrix is labeled by another such vector $F$,
\begin{equation}
\xi_F=\xi_1^{F_1}\xi_2^{F_2}\xi_3^{F_3}\xi_4^{F_4},
\end{equation}
with $\xi_\mu=\gamma_\mu^*$.
The two unitarily equivalent sets of matrices are then
\begin{eqnarray}
\overline{(\gamma_S\otimes \xi_F)}_{AB}
&\equiv& \frac14 {\rm Tr}\left[
\gamma_A^\dagger 
\gamma_S^{\vphantom{\dagger}}
\gamma_B^{\vphantom{\dagger}}
\gamma_F^\dagger \right],
\label{eq:singlebardef}
\\
\overline{\overline{(\gamma_S\otimes \xi_F)}}_{AB}
&\equiv& \sum_{CD}\frac{(-)^{A\cdot C}}{4}
\overline{(\gamma_S\otimes \xi_F)}_{CD}
\frac{(-)^{D\cdot B}}{4}.
\label{eq:doublebardef}
\end{eqnarray}
Using these relations one can trace the connection between
the $16^2$ choices for the indices $AB$ in the propagator
(\ref{eq:freeprop}) to the spin and taste indices in the
more familiar form $(\gamma_S\otimes \xi_F)$.

\subsection{Definition of the asqtad action}

The asqtad action is~\cite{asqtada,asqtadb,asqtadlepage} 
\begin{widetext}
\begin{eqnarray}
S_\text{asqtad} &=& 
\sum_{n} \bigg[ \bar{\chi}(n)
\sum_{\mu} \eta_{\mu}(n) \Big( 
\nabla_\mu^\text{F7L} \chi(n)
+ \frac{1}{8} [\nabla_\mu^\text{T1} - \nabla_\mu^\text{T3}] 
\chi(n) \Big) 
+ (m/u_0) \bar{\chi}(n)\chi(n) \bigg] \,,
\label{eq:asqtad}
\\
\nabla_\mu^\text{F7L} \chi(n) &=&\frac{1}{2}
[W_\mu (n) \chi(n + \hat{\mu}) - 
W^{\dagger}_{\mu}(n-\hat{\mu}) \chi(n-\hat{\mu})]\,,
\label{eq:fat7pluslepage}
\\
\nabla_\mu^\text{T1} \chi(n) &=&\frac{1}{2 u_0}
[ U_\mu(n) \chi(n + \hat{\mu}) - 
U^\dagger_\mu(n-\hat{\mu})\chi(n-\hat{\mu}) ]\,,
\\
\nabla_\mu^\text{T3} \chi(n) &=&\frac{1}{6 u_0^3}
[U(n,n+3\hat{\mu}) \chi(n + 3\hat{\mu}) - 
U(n,n-3\hat{\mu})\chi(n-3\hat{\mu})]\,,
\end{eqnarray}
\end{widetext}
where $W_\mu (n)$ is a smeared link constructed using the Fat7 blocking
transformation~\cite{asqtada,asqtadb} combined with Lepage's
prescription~\cite{asqtadlepage} and tadpole improvement~\cite{LM},
and $U(n,n\pm3\hat{\mu})$ are products of 3 thin links in the $\mu$
direction starting at position $n$.
Finally, $u_0$ is the tadpole improvement factor~\cite{LM}, which we
take to be the fourth-root of the average plaquette.

\section{Irreducible representations for covariant bilinears}
\label{app:irreps}

In this appendix we sketch the demonstration that the covariant
bilinears $\CO_{S\otimes F}^{\rm cov}$ of Eq.~(\ref{eq:Ocovb})
fall into the irreps listed in Table~\ref{tab:irreps}
under the lattice symmetry group. Although this result is
likely known to workers in the field, we have not found
a demonstration in the literature.
In particular, in their seminal work on staggered fermions,
Golterman and Smit described the full lattice group~\cite{GS},
but focused on constructing operators transforming
as irreps of the smaller timeslice group,
which classifies eigenstates of the transfer matrix~\cite{GSbaryons,Gmesons}.
Verstegen subsequently classified the irreps of 
bilinears living on a single $2^4$ hypercube~\cite{Verstegen}.
The symmetry group in this case is smaller than that for the
zero momentum covariant bilinears, since translations are
excluded. Thus, although Verstegen's work will be useful
in the following, the irreps he finds are in general smaller
than those for covariant bilinears.

Perturbative calculations of matching factors also
give information on the irreps, since operators living
in different irreps have different matching factors.
This information is, however, incomplete since results are
available only at finite order (usually 1-loop), and differences
could show up at higher order.

In the subsequent discussion we use the presentation of the
lattice group and method of analysis 
(as well as the notation) of Ref.~\cite{toolkit}.
We refer to this reference for most of the technical details.
An alternative approach is that of Ref.~\cite{Ggroup}.

For operators having zero physical momentum, 
the group of transformations is
\begin{eqnarray}
{\CG_0}&=&\Gamma_{4,1} \semitimes W_4
\label{eq:CG0}
\nonumber\\
\Gamma_{4,1}&=&\{\Xi_\mu,C_0\}\,,\qquad
W_4 = \{R_{\mu\nu},I_s\}\,.
\end{eqnarray}
Here $W_4$ is the hypercubic group generated by rotations
$R_{\mu\nu}$ and spatial inversion $I_s$,
while $\Gamma_{4,1}$ is the Clifford group in five-dimensional
Minkowski space generated by lattice charge conjugation, $C_0$,
and single-site translations $\Xi_\mu$.\footnote{%
In general $\Xi_\mu$ are single-site translations with the
momentum factor $e^{ip'_\mu}$ removed, but this removal
in not needed as $p'=0$.}
The symbol ``$\semitimes$'' indicates a semidirect product.
In the analysis of Verstegen the translations $\Xi_\mu$ are
absent, leaving only the group $W_4$ combined with $C_0$. 
Thus the constraints
he finds are weaker than those obtained from $\CG_0$.

Under translations the covariant bilinears pick up a sign
$(-)^{\tilde F_\mu}$, where $\tilde F_\mu=\sum_{\nu\ne\mu}F_\nu$.
This is shown in the following appendix. 
Similarly the bilinears have a definite parity under $C_0$
(which is straightforward to calculate but not needed in the following).
Thus the bilinears reside in 1-d irreps of $\Gamma_{4,1}$
characterized by five parities. In Ref.~\cite{toolkit} these are
called
\begin{equation}
\Delta^{(4,1)}(\pm,\pm,\pm,\pm,\xi_C)\,,
\end{equation}
where the first four arguments are the translation signs
under $\Xi_1$, $\Xi_2$, $\Xi_3$ and $\Xi_4$,
while the last is the parity under $C_0$.
Since $\CG_0$ is a semidirect product, one must,
for each irrep $\Delta^{(4,1)}$,
find the subgroup of $W_4$ which leaves the irrep invariant.
The bilinears are then classified into irreps of this ``little
group''. These induce representations of the
full group that are known to be irreducible.

The action of the rotations and spatial inversion
which form $W_4$ is discussed in the next appendix.
All we need to know here is that
both transformations act simultaneously on spin and taste indices.
Thus $O^{\rm cov}_{S\otimes F}$ is transformed, up to a sign, into
$O^{\rm cov}_{S_R\otimes F_R}$, where
$S_R$ and $F_R$ are the hypercube vectors obtained from $S$ by $F$
by the transformation under consideration.

We now begin the classification of bilinears into irreps.
For taste singlet bilinears, the $\Gamma_{4,1}$ irrep is
$\Delta^{(4,1)}(+,+,+,+,\xi_C)$ and the little group is
the full $W_4$~\cite{toolkit}.
The same little group holds for taste $\xi_5$ [$F=(1111)$]
for which the irrep is
$\Delta^{(4,1)}(-,-,-,-,\xi_C)$.
In both cases we can use the analysis of Verstegen, who shows
(see his Table 3 for irreps of the rotation subgroup, together
with the discussion in his Sec.~5 of how inversion combines
irreps) that each of the five types of spin lives in a single
irrep.\footnote{%
Verstegen's rotations and inversions are about the center of
the hypercube, rather than the standard choice of being
about a lattice point. These choices differ, however,
by translations, which, for the taste singlet and $\xi_5$ operators
are simply signs, and do not lead to changes in the dimensionality
of the resulting irreps.}
Explicitly, the irreps for taste singlets have spin-tastes
\begin{equation}
(I\otimes I),\
(\gamma_\mu\otimes I),\ 
(\gamma_{\mu\nu}\otimes I),\
(\gamma_{\mu5}\otimes I) \ \& \ 
(\gamma_5\otimes I).
\end{equation}
Here $\mu$ and $\nu$ run from $1-4$ except that $\mu<\nu$.
These are the five taste-singlet irreps appearing in Table~\ref{tab:irreps}.
The same five spins apply also to taste $\xi_5$.

Next we consider bilinears with taste $\xi_\mu$ and $\xi_{\mu5}$.
Choosing $\mu=4$ for definiteness, the $\Gamma_{4,1}$ irreps are
\begin{equation}
\Delta^{(4,1)}(-,-,-,+,\xi_C)
\ \ {\rm and}\ \
\Delta^{(4,1)}(+,+,+,-,\xi_C)\,,
\end{equation}
respectively.
In both cases the little group is $W_3\times Z_2$, with
$W_3$ the cubic group $\{R_{ij},I_s\}$ while $Z_2$ is
generated by the axis inversion symmetry in the 4'th 
direction~\cite{toolkit}.
Determining the transformations under rotations,
one finds that the following spins live in 3-d
irreps of $W_3$ (either the $1$ or the $\overline{1}$
in the notation of Ref.~\cite{Mandula}):
\begin{equation}
\gamma_j,\
\gamma_{j5},\
\gamma_{j4},\ {\rm and}\
\epsilon_{jkl}\gamma_{kl}
\qquad (j,k,l=1{-}3)\,.
\end{equation}
The remaining spins ($I$, $\gamma_5$, $\gamma_4$ and $\gamma_{45}$)
live in one of the two 1-d irreps.
Extending these irreps of the little group
to the full group by acting with
the ``missing'' generators, i.e.\ $R_{4k}$, their size is
multiplied by a factor of $4$, the dimension of the orbit of
the $\Gamma_{4,1}$ irrep under $W_4$.
Thus (choosing taste vector for definiteness) one ends up
with four 12-d irreps and four 4-d irreps:
\begin{align}
\begin{split}
&(\gamma_\mu\otimes\xi_\nu),\
(\gamma_{\mu5}\otimes \xi_\nu),\
(\gamma_{\mu\nu}\otimes \xi_\nu),\
(\gamma_{\mu\rho}\otimes \xi_\nu),
\\
& (I\otimes\xi_\nu),\
(\gamma_5\otimes\xi_\nu),\
(\gamma_\nu\otimes\xi_\nu),\
(\gamma_{\nu5}\otimes\xi_\nu),
\end{split}
\end{align}
where $\mu\ne\nu$, $\rho\ne\nu$ and $\mu<\rho$.
These are the eight taste-vector
irreps appearing in Table~\ref{tab:irreps}.
The same set of spins appear for the axial taste bilinears
(with $\xi_\mu\to \xi_{\mu5}$).

In this case, the results differ from those obtained for
single-hypercube bilinears. For example, Verstegen finds that
the spin-scalar, taste-vector bilinears split into two irreps,
a 1-d irrep $\sum_\mu (I\otimes\xi_\mu)$ and a 3-d irrep
consisting of the differences $(I\otimes\xi_\mu)-(I\otimes\xi_\nu)$.
For covariant bilinears, by contrast, one has a single
4-d irrep, $(I\otimes \xi_\mu)$.

Finally, we consider the taste tensors. If the taste is
$\xi_{12}$, the $\Gamma_{4,1}$ irrep is 
$\Delta^{(4,1)}(-,-,+,+,\xi_C)$.
The little group is $D_4\otimes D_4$, where the 
first dihedral group $D_4$ is generated by
$R_{12}$ and $I_1$ (the axis inversion operator
in the 1st direction), while the second $D_4$ is
generated by $R_{34}$ and $I_3$.
$D_4$ has four 1-d and one 2-d irreps.
The bilinears decompose into a single 4-d irrep
of $D_4\times D_4$ (spin $\gamma_j\gamma_k$, with $j=1,2$
and $k=3,4$), four 2-d irreps
(spins $\gamma_j$, $\gamma_{j5}$, $\gamma_k$ and $\gamma_{k5}$),
and four 1-d irreps
(spins $I$, $\gamma_5$, $\gamma_{12}$ and $\gamma_{34}$).
The orbit in this case is six dimensional, so the induced
irreps of $\CG_0$ are the 24 dimensional
\begin{equation}
(\gamma_{\mu\rho}\otimes\xi_{\mu\nu}), 
\end{equation}
with $\mu<\nu$, $\rho\ne\mu$ and $\rho\ne\nu$,
the four 12-d irreps
\begin{equation}
(\gamma_\mu\otimes\xi_{\mu\nu})\,\
(\gamma_{\mu5}\otimes\xi_{\mu\nu})\,\
(\gamma_\rho\otimes\xi_{\mu\nu})\, \ \& \
(\gamma_{\rho5}\otimes\xi_{mu\nu}),
\end{equation}
and the four 6-d irreps
\begin{equation}
(I\otimes\xi_{\mu\nu})\,\
(\gamma_{5}\otimes\xi_{\mu\nu})\,\
(\gamma_{\mu\nu}\otimes\xi_{\mu\nu})\, \ \& \
(\gamma_{\rho\sigma}\otimes\xi_{\mu\nu}),
\end{equation}
with indices constrained as above together with
$\sigma$ differing from $\mu$, $\nu$ and $\rho$.
Altogether, these are the nine taste-tensor irreps appearing in
Table~\ref{tab:irreps}. 

\section{Symmetry constraints on propagator and vertices}
\label{app:symm}

In this appendix we describe how lattice translation symmetry
constrains the form of the quark propagator and the vertices
of covariant bilinears.

The fermion fields transform under translations as~\cite{GS}
\begin{equation}
\chi(n) \to \zeta_\mu(n)\chi(n\!+\!\hat\mu)
\ {\rm and}\ \,
\bar\chi(n) \to \zeta_\mu(n)\bar\chi(n\!+\!\hat\mu)\,.
\end{equation}
The translation phases can be chosen to be 
\begin{equation}
\zeta_\mu(n) = (-)^{\sum_{\nu>\mu} n_\nu}
= (-)^{n_\zeta \cdot\hat\mu} = (-)^{n \cdot \hat\mu_\eta}
\,,
\end{equation}
where
\begin{equation}
n_\zeta = (n_2+n_3+n_4,n_3+n_4,n_4,0)
\end{equation}
and
\begin{equation}
n_\eta=  (0, n_1, n_1+n_2, n_1+n_2+n_3),
\end{equation}
and we have used the identity
\begin{equation}
n_\zeta \cdot m = n \cdot m_\eta
\,.
\end{equation}
Thus the momentum-space field (\ref{eq:goodmomfield}) transforms as
\begin{eqnarray}
\chi_A(p') &\!\!\to\!\!& 
\sum_n e^{-ip'\cdot n} (-)^{A \cdot n} (-)^{n\cdot \hat\mu_\eta}
\chi(n\!+\!\hat\mu)
\\
&\!\!=\!\!& e^{ip'_\mu} (-)^{A_\mu} \sum_{m} e^{-i p'\cdot m}
(-)^{m\cdot (A\!+\!\hat\mu_\eta)} \chi(m)
\\
&\!\!=\!\!& e^{ip'_\mu} (-)^{A_\mu} \delta_{A+_2\hat\mu_\eta,C} \phi_C(p')
\\
&\!\!=\!\!& e^{ip'_\mu} \ixibar{\mu}_{AC} \phi_C(p')\,,
\end{eqnarray}
where $+_2$ indicates addition mod 2.
In the last step we have used
\begin{equation}
\ixibar{\mu}_{AC} = (-)^{A_\mu} \delta_{A+_2\hat\mu_\eta,C} 
\,.
\end{equation}
One can similarly show that
\begin{equation}
\bar\phi_B(-q') \to e^{-iq'_\mu} \bar\phi_D(-q') 
\ixibar{\mu}_{DB}
\,.
\end{equation}
These results show explicitly how translations by a single site correspond
(once momentum factors are removed) to taste rotations~\cite{GS}.

Using the translation invariance of the action, one thus learns
that the momentum space propagator (\ref{eq:Sdef}) satisfies
\begin{eqnarray}
\lefteqn{S(p')_{AB} N_{\rm site}\equiv 
\langle \phi_A(p') \bar\phi_B(-p')\rangle}
\nonumber \\
&=& \ixibar{\mu}_{AC} 
    \langle\phi_C(p')\bar\phi_D(-p') \rangle 
    \ixibar{\mu}_{DB}
\end{eqnarray}
from which the result (\ref{eq:translateS}) follows.
As explained in the main text, it follows that the propagator
is taste-singlet.

We now turn to the implications of translation invariance for the
(unamputated) vertex, Eq.~(\ref{eq:Lambdadef}).
As for the propagator, translating the external fields
lead to multiplications by $\ixibar{\mu}$, as well as to phase factors
which cancel in our kinematics.
To determine the effect of
translations on the bilinear operator (\ref{eq:Ocovb}) 
we first note that the $\bar\chi$ and $\chi$ fields together lead to
the sign $(-)^{(S-F)_\zeta\cdot\hat\mu}$.
Combining this with the sign resulting from
translating the phases in the operator,
\begin{eqnarray}
\lefteqn{
\singlebar{S}{F}_{x+\hat\mu,x+\hat\mu+S-F}
= }
\nonumber\\
&&(-)^{\widetilde F_\mu} (-)^{(S-F)_\zeta\cdot\hat\mu}
\singlebar{S}{F}_{x,x+S-F},
\end{eqnarray}
we find
(dropping flavor indices for clarity)
\begin{eqnarray}
\CO_{S\otimes F}^{cov}
&\to& 
(-)^{\widetilde F_\mu}\
\CO_{S\otimes F}^{cov}
\,.
\label{eq:covariance}
\end{eqnarray}

Combining these results we see that the
vertex functions satisfy, for each $\mu$,
\begin{equation}
\Lambda^{S\otimes F}(p') = (-)^{\widetilde F_\mu}\
\ixibar{\mu}\ \Lambda^{S\otimes F}(p')\ \ixibar{\mu}
\,.
\label{eq:translateLambdares}
\end{equation}
This implies that $\Lambda(p)$ must have taste $F$,
because
\begin{equation}
\ixibar{\mu}\ \doublebar{S}{F}\ \ixibar{\mu}
= (-)^{\widetilde F_\mu}\; \doublebar{S}{F}
\,.
\end{equation}
When we amputate the vertex using inverse propagators,
which we know, from above, are taste singlets, the resulting
amputated vertex will also have taste $F$.
This shows that, if one uses covariant bilinears,
there can be no mixing with other tastes.

We next discuss the constraints due to spatial inversion
symmetry, $I_s$. This acts on the fields as
\begin{equation}
\chi(n) \to \eta_4(n) \chi(n_S)\,,\
\bar\chi(n) \to \eta_4(n) \bar\chi(n_S)\,,
\end{equation}
where $n_S=I_s^{-1}n$. By manipulations analogous to
those given above, one can rewrite these transformations
in terms of the momentum-space fields, finding:
\begin{equation}
\phi(p') \to \doublebar{4}{4} \phi(p'_S)
\,,\
\bar\phi(-p') \to \phi(-p'_S) \doublebar{4}{4}
\,.
\end{equation}
From the invariance of the action under $I_s$ it 
follows that
\begin{align}
\begin{split}
S(p') &= \doublebar{4}{4} S(p'_S) \doublebar{4}{4}
\\ 
\Rightarrow&\
S^{-1}(p') = \doublebar{4}{4} S^{-1}(p'_S) \doublebar{4}{4}
\,.
\label{eq:inversiononS}
\end{split}
\end{align}
The conjugation by $\doublebar{4}{4}$ flips the sign of
each spatial component of all spin and taste matrices.
Since we know, however, that $S^{-1}$ is a taste singlet,
the effect of the conjugation is to
replace each $\gamma_\mu$ with its spatial inverse.
The relation (\ref{eq:inversiononS}) thus has exactly
the same implication as the corresponding result 
for fermions without the taste degree of freedom.
As noted in the main text, combined with rotations,
one finds that $S^{-1}$ has the form given in
Eq.~(\ref{eq:Sinvform}).
The appearance of odd powers of $p'$ in this result is
due to the spatial inversion symmetry.

For the unamputated vertex one finds that inversion
symmetry leads to 
\begin{equation}
\Lambda_{S\otimes F}(p') = \eta_4(\Delta)
\doublebar{4}{4}\; \Lambda_{S\otimes F}(p'_S)\; \doublebar{4}{4}\,,
\label{eq:inversiononLambda}
\end{equation}
where $\Delta=S-F$.
Multiplying from left and right with $S^{-1}(p')$ and
using the relation (\ref{eq:inversiononS}), one can
convert this into a result of the same form for the amputated vertex:
\begin{equation}
\Gamma_{S\otimes F}(p') = \eta_4(\Delta)
\doublebar{4}{4} \;\Gamma_{S\otimes F}(p'_S)\; \doublebar{4}{4}\,.
\label{eq:inversiononGamma}
\end{equation}
We consider only the consequences of this result for the
momentum independent part of the vertex, i.e.\ that which
survives in the continuum limit (when multiplied by an
appropriate matching factor). Then $\Gamma_{S\otimes F}$
is simply a $16\times16$ matrix, having taste $F$ (from translation
invariance) but as yet undetermined spin:
\begin{equation}
\Gamma_{S\otimes F} = \sum_{S'} c_{SS'}^F \doublebar{S'}{F}
\,.
\label{eq:Gammaforminversion}
\end{equation}
The constraint (\ref{eq:inversiononGamma}) implies that
the only non-vanishing constants, $c_{SS'}^F$ are those for
which $S'$ satisfies $\eta_4(S'-S)=1$. This is because
\begin{align}
\begin{split}
\doublebar{4}{4}\ \doublebar{S'}{F}\ &\doublebar{4}{4} 
\\
&= \eta_4(S'-F) \doublebar{S'}{F}
\end{split}
\end{align}
and
\begin{equation}
\eta_4(S'-F) = \eta_4(S'-S)\eta_4(\Delta)
\,.
\end{equation}
Thus inversions alone allow several choices of $S'$,
those satisfying $S'_1+S'_2+S'_3=_2S_1+S_2+S_3$.

To further constrain the propagator and vertices we
turn to the final discrete symmetry, namely rotations.
Here the analysis is more involved, since rotations
mix bilinears.
Consider the $(\mu\nu)$ rotation generator defined such
that 
\begin{equation}
p'_R=R^{-1} p'\,, 
(p'_R)_\mu = -p'_\nu\,,
(p'_R)_\nu = p'_\nu\,,
(p'_R)_\rho = p'_\rho\,,
\end{equation}
where $\mu$, $\nu$ and $\rho$ are all different.
We find that the inverse propagator satisfies
\begin{equation}
S^{-1}(p') = {\cal R}\; S^{-1}(p'_R)\; {\cal R}^{-1}
\label{eq:Sinvrots}
\end{equation}
where
\begin{equation}
{\cal R} = \frac12 
\overline{\overline{([I+\gamma_{\mu\nu}]\otimes [I+\xi_{\mu\nu}])}}
\,.
\end{equation}
The key property of ${\cal R}$ is that it rotates the 
spin and taste matrices, e.g.\
\begin{align}
{\cal R} \overline{\overline{(\gamma_\mu\otimes I)}} {\cal R}^{-1}
&= -  \overline{\overline{(\gamma_\nu\otimes I)}}
\,,\\
{\cal R} \overline{\overline{(\gamma_\nu\otimes I)}} {\cal R}^{-1}
&=   \overline{\overline{(\gamma_\mu\otimes I)}}
\,,\\
{\cal R} \overline{\overline{(I\otimes\xi_\mu)}} {\cal R}^{-1}
&= -  \overline{\overline{(I\otimes\xi_\nu)}}
\,.
\end{align}
The result (\ref{eq:Sinvrots}) is the final input which leads to the
general form of the propagator, Eq.~(\ref{eq:Sinvform}).
Given that $S^{-1}$ is a taste-singlet, 
(\ref{eq:Sinvrots}) enforces that each $\gamma_\mu$
must be multiplied by a power of $p_\mu$.

The implication of rotation invariance for amputated
vertices is
\begin{align}
\Gamma_{S\otimes F}(p')
&= \psi(S,F)\; {\cal R}\; \Gamma_{S_R\otimes F_R}(p'_R)\; {\cal R}^{-1}
\label{eq:rotatevertex}
\\
\psi(S,F) &=
\frac1{16}\tr\left[
{\cal R}\doublebar{S_R}{F_R}{\cal R}^{-1}\doublebar{S}{F}^\dagger\right]
\,.
\end{align}
Note that, unlike for translations and spatial inversion, the
vertices on the two sides of (\ref{eq:rotatevertex})
involve different operators. This is as
expected since the operators fall into non-trivial irreps 
under the full lattice group.
Note that the signs $\psi(S,F)$ are such that 
(\ref{eq:rotatevertex}) is satisfied if
$\Gamma_{S\otimes F}(p')=\doublebar{S}{F}$.
We will show that, up to a constant,
this is the only momentum-independent
solution to (\ref{eq:rotatevertex})
that is also consistent with the relations from
translations and spatial inversion.

Indeed, from translations and spatial inversion we know the form 
of the momentum-independent part of the amputated vertices to be
that of Eq.~(\ref{eq:Gammaforminversion}).
Applying (\ref{eq:rotatevertex}) we learn that the
constants satisfy
\begin{equation}
c_{SS'}^F = c_{S_R S_R'}^{F_R} \frac{\psi(S,F)}{\psi(S',F)}
\,.
\label{eq:relbetweenconsts}
\end{equation}
At first sight, this appears to simply relate the constants
appearing in the expansions of $\Gamma_{S\otimes F}$
and $\Gamma_{S_R\otimes F_R}$.
However, if we apply (\ref{eq:relbetweenconsts}) twice we
obtain
\begin{equation}
c_{SS'}^F=c_{SS'}^F \frac{\psi(S,F)\psi(S_R,F_R)}
                         {\psi(S',F)\psi(S'_R,F_R)}
\,.
\label{eq:rel2}
\end{equation}
Here we have used the result that $(S_R)_R=S$ for hypercube
vectors, since their elements are binary numbers.
It is straightforward to show that
\begin{equation}
\psi(S,F)\psi(S_R,F_R) = (-)^{S_\mu+S_\nu} (-)^{F_\mu +F_\nu}
\,,
\end{equation}
so that (\ref{eq:rel2}) becomes
\begin{equation}
c_{SS'}^F=c_{SS'}^F (-)^{S_\mu+S_\nu+S'_\mu+S'_\nu}
\,.
\end{equation}
Thus we learn that the only non-vanishing constants
are those for which $S'_\mu+S'_\nu=_2S_\mu+S_\nu$ for
all pairs $(\mu,\nu)$. The only solutions are
$S'=S$ and $S'=_2 S + (1111)$.
This ambiguity is expected, since rotations
alone allow mixing, e.g.\ between $\gambar{\mu}$ and
$\gambar{\mu5}$.
However, if we also enforce spatial inversion invariance,
which, as seen above, implies $\eta_4(S')=\eta_4(S)$,
then we find that only $S'=S$ is allowed.
Thus we finally attain the desired result that
\begin{equation}
\Gamma_{S\otimes F}(p') \propto \doublebar{S}{F} + O(a)
\,,
\end{equation}
where the $O(a)$ indicates momentum and mass dependent terms.

For completeness we note that one can obtain covariant operators
containing derivatives by adding appropriate signs in the
sum over $\Delta$ in Eq.~(\ref{eq:Ocovb}). In particular,
if the derivative is in the $\mu$'th direction,
instead of adding the two terms in (\ref{eq:symmsum}), 
one takes the difference when $\Delta_\mu=1$.
This leads to the correspondence
\begin{eqnarray}
\lefteqn{
\frac1V a^3 \int d^4x\; \bar Q\partial_\mu (\gamma_S\otimes\xi_F) Q
\simeq}
\nonumber\\
&& \frac{1}{N_{\rm site}} \sum_n 
\frac{1}{N_\Delta} \sum_{|\Delta|=|S\!-\!F|}  \Delta_\mu
\nonumber\\
&&\ \ 
\bar\chi(n) \singlebar{S}{F}_{n,n\!+\!S\!-\!F} \;
\CU_{n,n\!+\!\Delta}\; \chi(n\!+\!\Delta)
\,.
\end{eqnarray}
The only difference from (\ref{eq:Ocovb}) is the factor of $\Delta_\mu$.
This construction only works if $\Delta_\mu\ne 0$, i.e.\ if the spin-taste
of the operator is such that the $\bar\chi$ and $\chi$ fields are 
already separated in the $\mu$'th direction.
If they are not, one must use a two-step difference to get 
an operator containing a derivative~\cite{SP}.

\section{Perturbative matching for covariant bilinears}
\label{app:pert}

In this appendix we describe briefly how the use of covariant bilinears
changes the one-loop matching factors compared to those
for hypercube bilinears. The latter have been calculated
for our choices of fermion and gauge action in Ref.~\cite{KLS},
following the earlier work of Refs.~\cite{GS,DS,PS,LS}.
We also present numerical results for our choices of action,
since these are not given in Ref.~\cite{KLS}

It is instructive to compare the tree-level matrix elements of
the hypercubic and covariant bilinears
between external quark ``states'' with physical
momenta $p'+\pi C$ (outgoing from $\bar\chi$) and $p'+\pi D$
(incoming to $\chi$). As explained in Ref.~\cite{PS}, the
matrix element of a hypercubic operator is
\begin{eqnarray}
\lefteqn{M(S\otimes F;{\rm hyp})_{CD}^{(0)} =}
\nonumber\\
&&\sum_{MN} E_M(p')E_N(-p') 
\doublebar{MSN}{MFN}_{CD}
\,,
\label{eq:Mhyp}
\end{eqnarray}
where, like $S$ and $F$, $M$ and $N$ are hypercube vectors.
(Note that ``hyp'' indicates hypercubic operator and should
not be confused with ``HYP'' for HYP-smearing.)
The functions which enter are
\begin{equation}
E_M(k) = \prod_\mu \frac12\left(
e^{-ik_\mu/2}+ (-)^{\widetilde M_\mu} e^{i k_\mu/2} \right),
\end{equation}
which are thus products of cosines or sines for the different components.
We see from (\ref{eq:Mhyp}) that, even in this tree-level
matrix element, all combinations of spin
and taste appear which satisfy $S'-F'=_2 S-F$, i.e.\ which have
the same number of links.
This mixing is, however, suppressed by powers of
$a$, since if $M\ne0$ then $\widetilde M\ne0$, and there is
at least one factor of $\sin(p'_\mu/2)\propto a {p'_\mu}^{\rm phys}$
in $E_M$. These factors of $a$ correspond to the fact that the
hypercube operators, when written in terms of irreps of the
translation group, break up into the desired dimension-3 bilinear
plus additional dimension-4 and higher operators containing derivatives.

If one projects out the part of this vertex with the same spin and taste
as the initial bilinear (as one does in NPR), one finds
\begin{eqnarray}
\lefteqn{
\frac1{48}\Tr\left[\doublebar{S}{F}^\dagger M(S\otimes F;{\rm hyp})^{(0)} 
\right] }
\nonumber\\
&=& \sum_{M} E_M(p') E_M(-p') (-)^{(S-F)\cdot \widetilde M}
\\
&=& \prod_\mu \cos[p'_\mu (S-F)_\mu] \equiv V_{S\otimes F}(p')
\,.
\label{eq:vertexfunction}
\end{eqnarray}
To obtain the last line we have used
the sum rule given in Eq.~(A8) of Ref.~\cite{LS}.
Thus the tree-level kinematic factor associated with the hypercube bilinear
is the vertex factor $V_{S\otimes F}$. This factor necessarily
tends to unity in the continuum limit, but the $O(a^2)$ corrections
contained in the cosines can be significant in practice.

At one-loop level, some of the hypercube operators mix. This mixing
arises from the so-called X-diagrams (see, e.g., Ref.~\cite{KLS} for
a figure explaining this terminology), in which the momentum flowing
through the bilinear is not of $O(a)$ but rather of $O(1)$ (since it
is inside a loop integral). This mixing is not suppressed by powers of $a$.

Now consider the covariant bilinears of Eq.~(\ref{eq:Ocovb}). 
A key point is that the sign arising from 
the matrix $\singlebar{S}{F}$ is independent
of $\Delta$. Thus the sum over $\Delta$ can be done, and leads exactly
to the vertex factor $V_{S\otimes F}(p')$. The tree-level vertex is simply
\begin{equation}
M(S\otimes F;{\rm cov})_{CD}^{(0)} =
V_{S\otimes F}(p') \doublebar{S}{F}_{CD} 
\,,
\label{eq:Covvertex}
\end{equation}
with no mixing with other spins and tastes.
Because of the lack of mixing, one can read off the kinematical factor
associated with this vertex without the need for projection.
The result is that the kinematical factor is the {\em same} as that for
hypercube operators.

Since there is no mixing in the vertex (\ref{eq:Covvertex}),
irrespective of the value of $p'$, we expect that there will be
no mixing between covariant bilinears in the one-loop calculation.
This is indeed what we find by explicit calculation.
Furthermore, it turns out that the diagonal (non-mixing) parts of
the matching factors are identical to those for hypercube bilinears.
For the X-diagrams, this is because the same
factor $V_{S\otimes F}$ occurs in both tree-level vertices.
For the ``Y-diagrams'' (those involving a gluon coupling to the
vertex---see, e.g., Ref.~\cite{KLS}) the reason for the equality
is similar. The remaining diagrams (self-energy and tadpoles) are
the same for both operators.

Thus we arrive at a very simple result. At one-loop order, the diagonal
matching of covariant operators is identical to that for hypercube operators,
while the off-diagonal matching coefficients vanish.
We stress that the equality of diagonal matching factors
should not hold at higher orders in perturbation theory. 
One way to see this is that the hypercube operators with 
different spin-taste that arise due to one-loop mixing 
can mix back with the original operators at two-loop order. 
Such contributions are not present for the covariant operators.

The rest of this appendix is devoted to providing
numerical results for bilinear one-loop matching factors
for the Symanzik gauge action and our choices of valence
fermions and links in the bilinears.
Analytic formulae are given in Ref.~\cite{KLS}, 
but that work quotes numerical values for several
choices of fermion actions and operators differing from
those we use.
In particular, when HYP-smearing
we use the HYP(1) choice of smearing parameters.

We are ultimately interested in perturbative predictions for
the matching factors $Z_{S\otimes F}$ relating operators
in the ``lattice scheme'' 
(i.e.\ the bare operators we place on the lattice and
use in simulations)
to those in the RI$'$ scheme.
These are the matching factors
we obtain non-perturbatively in our simulations
using Eq.~(\ref{eq:NPRvertex}).
However, perturbative calculations typically give results for
matching from the lattice scheme to an intermediate continuum scheme,
usually $\overline{\rm MS}$.
Thus to obtain the full matching factors one must determine
the matching between $\overline{\rm MS}$ and RI$'$ schemes. 
This latter matching can be done in the continuum. 

These considerations lead to the ``master formula''\footnote{%
For the sake of clarity,
we have made the dependence on $a$ explicit on the left-hand side,
although this is left implicit in the main text.}
\begin{eqnarray}
Z_{\CO}^{{\rm RI}',{\rm LAT}}(\mu,a)
&=&
\exp\left[-\int_{\lambda(\mu_0)}^{\lambda(\mu)} d\lambda 
\frac{\gamma_{\CO}(\lambda)}{\beta(\lambda)}\right] 
\nonumber\\
& \times& Z_{\CO}^{{\rm RI}',\overline{\rm MS}}(\mu_0)
\times Z_{\CO}^{\overline{\rm MS},{\rm LAT}} (\mu_0,a)
\,. \label{eq:Zmaster}
\end{eqnarray}
Here $\lambda=\alpha/(4\pi)$,
with $\alpha$ always evaluated in the $\overline{\rm MS}$ scheme.
$\gamma_{\CO}$ is the anomalous dimension of the operator
($\CO$ is shorthand for $S\otimes F$),
and $\beta(\lambda)$ the $\beta$-function.
In words, this equation says that one way of matching from the RI$'$ scheme
at scale\footnote{%
In the main text this scale is denoted $p'$, but this symbol is used for
a dimensionless lattice momentum earlier in this appendix, so we
use $\mu$ here to denote a dimensionful energy scale.}
$\mu$ to the lattice scheme with spacing $a$ is to first
run in the RI$'$ scheme to an intermediate scale $\mu_0\approx 1/a$,
then convert to the $\overline{\rm MS}$ scheme at that scale,
and finally convert to the lattice scheme at scale $1/a$.
This formula allows one to have large values of the ratio
$\mu/\mu_0$, with the first factor on the right-hand side
summing the appropriate logarithms.

The one-loop results for matching from the lattice to the
$\overline{\rm MS}$ scheme have the form
\begin{eqnarray}
\lefteqn{Z_{\CO}^{\overline{\rm MS},{\rm LAT}}(\mu_0,a)
= \tilde u_0^{N_u}}
\nonumber\\
&&\left\{1 + \frac{\alpha(\mu_0)}{4\pi}
\left[-2\gamma_{\CO}^{(0)} \log(\mu_0 a)
+ C^{\overline{\rm MS}}_\CO - C^{\rm LAT}_{\CO}\right]\right\}
\,,
\label{eq:ZMSLAT}
\end{eqnarray}
$\gamma_\CO^{(0)}$ is one-loop anomalous dimension of the bilinear
(defined precisely in the following appendix)
and the $C$ are finite constants.
The continuum constants
are $C_I^{\overline{\rm MS}}=C_P^{\overline{\rm MS}}=10/3$,
$C_V^{\overline{\rm MS}}=C_A^{\overline{\rm MS}}=0$,
and
$C_T^{\overline{\rm MS}}=2/3$, and do not depend on the taste.
The factor of $\tilde u_0^{N_u}$ arises from possible mean-field
improvement. This will be discussed below, including the appropriate
values of $N_u$.
Without such improvement, $\tilde u_0=1$.
We stress again that
one should choose $\mu_0 \approx 1/a$ when using this result;
extending to other values of $\mu_0$ requires resumming the
leading logarithms using Eq.~(\ref{eq:Zmaster}).

A very important feature of the results (\ref{eq:Zmaster})
and (\ref{eq:ZMSLAT}) is that the anomalous dimensions
depend only on the spin $S$ but not on the taste $F$.
The same is true for $Z_{\CO}^{{\rm RI}',\overline{\rm MS}}$,
and, as seen above, the $C_{\cal O}^{\overline{\rm MS}}$.
This implies that if one takes ratios of matching 
factors having different tastes but the same spin, then
most of the terms in Eq.~(\ref{eq:Zmaster}) will cancel,
yielding
\begin{eqnarray}
\frac{Z_{S\otimes F}(\mu,a)}{Z_{S\otimes I}(\mu,a)}
&=&
\frac{Z_{S\otimes F}^{\overline{\rm MS},{\rm LAT}}(\mu_0,a)}
     {Z_{S\otimes I}^{\overline{\rm MS},{\rm LAT}}(\mu_0,a)}
\\
&=&
\tilde u_0^{|S|-|S-F|}
\left[
1 + \frac{\alpha(\mu_0)}{4\pi}
\delta^{S\otimes F}_{S\otimes I}
\right]
\,,
\label{eq:Zratsrelation}
\\
\delta^{S\otimes F}_{S\otimes I}
&=& 
C^{\rm LAT}_{S\otimes F}
- C^{\rm LAT}_{S\otimes 1}\,.
\label{eq:deltadef}
\end{eqnarray}
Here we have taken the denominators (arbitrarily) to be taste singlets.
The first line shows the cancellation of all except the
lattice to $\overline{\rm MS}$ matching factors,
and has the important consequence that the ratios are predicted
to be independent of $\mu$. This holds to all orders in PT,
and arises simply because it is only for momenta near the
lattice cut-off $1/a$ that taste dependence enters.
The lack of dependence on $\mu$ need not hold, however, for discretization
errors, so the ratios can depend on powers of $(a\mu)^2$.

The second line of Eq.~(\ref{eq:Zratsrelation}) gives
the one-loop result for the ratios, which, as shown in the third line,
depends only on the (difference of the) finite lattice
constants $C^{\rm LAT}$.
The values of these constants depend on whether mean-field improvement
(along the lines of Ref.~\cite{LM}) has been implemented.
In ratios, mean-field improvement amounts to dividing the links
in the bilinears by the fourth root of the plaquette
built from those links, $\tilde u_0$.
It is expected (and found) that bilinears with such rescaled links will
have better behaved perturbative expansions~\cite{LM}.
Since we have not implemented this rescaling in our non-perturbative
simulations, we must multiply by the rescaled bilinear
by $\tilde u_0$ raised to the power of the number of links in the
bilinear. These powers involve the length of
the hypercube vectors $S$ and $S-F$, 
where, e.g., $|S|=\sum_\mu |S_\mu|$.
Although it might appear that multiplying and dividing by the
same factors of $\tilde u_0$ would lead to no change, this is not
the case because for the external factors we use the non-perturbatively
determined value, while the impact of mean-field improvement
in the differences $\delta$ is 
evaluated in one-loop perturbation theory.
In effect, we are summing certain classes of diagrams to all orders
in PT by using the non-perturbative $\tilde u_0$.

We present results for the taste singlet constants
$C^{\rm LAT}_{S\otimes F1}$ and the differences
$\delta^{S\otimes F1}_{S\otimes I}$ in 
Tables~\ref{tab:CLAT}, \ref{tab:deltaS}, \ref{tab:deltaV} and \ref{tab:deltaT}
for the following choices of action and operators.
In all cases the gauge action is the tree-level 
improved Symanzik action.\footnote{%
The numerical results for cases (a) and (d) are directly obtained
from those in Ref.~\cite{KLS}, while those for cases (b), (c) and
(e) are new. The latter is new because Ref.~\cite{KLS}
did not consider mean-field
improvement of the asqtad bilinears.}
\begin{enumerate}
\item[(a)]
Mean-field improved naive staggered fermions 
with operators containing mean-field improved thin links.
In this case, $\tilde u_0$ is determined from the thin link plaquette,
and equals the $u_0$ discussed in the main text.
Mean-field improvement of the links replaces
$U_\mu$ with $U_\mu/\tilde u_0$. Mean-field improvement
of the action follows the prescription explained, 
for the present context, in Refs.\cite{PS,LS,KLS}.
The improvement of the action has no impact on the
differences $\delta$, but does change the constants $C$,
because the power of $\tilde u_0$ in Eq.~(\ref{eq:ZMSLAT}) becomes
$N_u=1-|S-F|$.
\item[(b)]
HYP-smeared staggered fermions with operators containing
HYP-smeared links. No mean-field improvement is
used, so that $\tilde u_0=1$. 
As noted above, HYP(1) smearing is used.
\item[(c)]
As in (b), except with mean-field improved HYP-smeared links,
with $\tilde u_0$ being the fourth root of the average plaquette
composed of HYP-smeared links. The action is also mean-field
improved, so that $N_u=1-|S-F|$ as in case (a).
\item[(d)]
Asqtad fermions with operators containing smeared
(``Fat7 $+$ Lepage'') links. No mean-field improvement.
\item[(e)]
As in (d), but with mean-field improvement of the 
links in the operators, $\tilde u_0$ now being the fourth root
of the plaquette composed of the
same smeared links as used in the operators.
Note that the asqtad action already includes some
tadpole improvement, and no further improvement is made to the action.
This means that $N_u=-|S-F|$ in Eq.~(\ref{eq:ZMSLAT}).
\end{enumerate}
We present results only for scalar, vector and tensor
bilinears, since multiplication of the operators 
by $\gamma_5\otimes\xi_5$ leaves the constants unchanged.
Thus those for pseudoscalars can be obtained from the 
results for scalars,  and 
results from axial bilinears from those for vectors.
In addition, three pairs of
tensor matching factors are equal, 
as displayed in Table~\ref{tab:deltaT}.

\begin{ruledtabular}
\begin{table}[tb!]
\begin{center}
\begin{tabular}{lrrrrr}
Spin (S)  & (a) & (b) & (c) & (d) & (e) \\
\hline 
$I$               & 34.12 & 3.29 & 2.71 & 4.83 & 4.83 \\
$\gamma_\mu$      & 0     & 0    & 0    &-1.91 &-6.57 \\
$\gamma_{\mu\nu}$ & -1.54 &-1.53 &-0.96 & 0.23 &-9.08 \\
\end{tabular}
\end{center}
\caption{Results for $C_{S\otimes I}^{\rm LAT}$
for the five choices of fermion action and operators explained in the text:
(a) Naive with mean-field improvement,
(b) HYP-smeared,
(c) HYP-smeared with mean-field improvement,
(d) asqtad with smeared links in operators,
(e) asqtad with smeared and mean-field improved links in operators.
The indices $\mu$ and $\nu$ are different.}
\label{tab:CLAT}
\end{table}
\end{ruledtabular}

\begin{ruledtabular}
\begin{table}[tb!]
\begin{center}
\begin{tabular}{lcrrrrr}
Taste (F) &links & (a) & (b) & (c) & (d) & (e) \\
\hline 
$\xi_\mu$      &1& 21.84 & 2.80 & 2.23 & 1.99 & 6.64 \\
$\xi_{\mu\nu}$ &2& 32.02 & 5.32 & 4.17 & 1.58 &10.88 \\
$\xi_{\mu5}$   &3& 37.41 & 7.66 & 5.93 & 0.21 &14.16 \\
$\xi_5$        &4& 41.52 & 9.90 & 7.59 &-1.51 &17.10 \\
\end{tabular}
\end{center}
\caption{Results for $\delta^{I\otimes F}_{I\otimes I}$, i.e.\
the finite coefficients for ratios involving scalar bilinears. 
The column ``links'' gives the number of links in the
operator with the given taste.
Choices of action and operators are
as in Table~\ref{tab:CLAT}.}
\label{tab:deltaS}
\end{table}
\end{ruledtabular}

\begin{ruledtabular}
\begin{table}[tb!]
\begin{center}
\begin{tabular}{lcrrrrr}
Taste (F)  & links & (a) & (b) & (c) & (d) & (e) \\
\hline 
$\xi_\mu$        &0& -5.32 & -1.05 &-0.48 & 3.26 &-1.40 \\
$\xi_{\mu\nu}$   &1& -3.46 &  0.32 & 0.32 & 0.09 & 0.09 \\
$\xi_\nu$        &2&  0.40 &  1.47 & 0.89 &-3.23 & 1.42 \\
$\xi_{\nu5}$     &2&  0.51 &  1.83 & 1.25 &-2.65 & 2.00 \\
$\xi_{\nu\rho}$  &3&  3.06 &  3.04 & 1.88 &-6.03 & 3.27 \\
$\xi_{5}$        &3&  3.44 &  3.38 & 2.23 &-5.49 & 3.82 \\
$\xi_{\mu5}$     &4&  5.80 &  4.64 & 2.91 &-8.78 & 5.18 \\
\end{tabular}
\end{center}
\caption{Results for $\delta^{\mu\otimes F}_{\mu\otimes I}$,
i.e.\ 
the finite coefficients for ratios involving vector bilinears. 
Notation as in Table~\ref{tab:CLAT}.
The indices $\mu$, $\nu$ and $\rho$ are all different.}
\label{tab:deltaV}
\end{table}
\end{ruledtabular}

\begin{ruledtabular}
\begin{table}[tb!]
\begin{center}
\begin{tabular}{lcrrrrr}
Taste (F)  & links & (a) & (b) & (c) & (d) & (e) \\
\hline 
$\xi_{\mu\nu}$          &0& 2.74 &-1.85 &-0.69 & 8.73 &-0.58 \\
$\xi_\mu$, $\xi_{\rho5}$&1&-0.91 &-1.05 &-0.47 & 3.82 &-0.84 \\
$\xi_{\mu\rho}$         &2&-0.36 &-0.02 &-0.02 &-0.09 &-0.09 \\
$\xi_5$                 &2& 0    & 0    & 0    & 0    & 0    \\
$\xi_\rho$, $\xi_{\mu5}$&3& 1.57 & 1.17 & 0.59 &-3.54 & 1.11 \\
$\xi_{\rho\sigma}$      &4& 3.72 & 2.45 & 1.30 &-6.82 & 2.49 \\
\end{tabular}
\end{center}
\caption{Results for 
$\delta^{\mu\nu\otimes F}_{\mu\nu\otimes I}$,
i.e.\ 
the finite coefficients for ratios involving tensor bilinears. 
Notation as in Table~\ref{tab:CLAT}.
The indices $\mu$, $\nu$, $\rho$ and $\sigma$ are all different.}
\label{tab:deltaT}
\end{table}
\end{ruledtabular}

For completeness, we also give the expressions for 
$\tilde u_0$ in PT. For HYP(1) smearing, we find
\begin{equation}
\tilde u_0^{\rm HYP} = 1 - C_F \frac{\alpha}{4\pi} 0.4331\,,
\label{eq:u0HYPinPT}
\end{equation}
where $C_f=4/3$, while for ``Fat7 $+$ Lepage'' smearing
\begin{equation}
\tilde u_0^{\rm ASQ} = 1 + C_F \frac{\alpha}{4\pi} 3.4897\,.
\label{eq:u0ASQinPT}
\end{equation}

We comment briefly on the values of the constants.
$C_{\gamma_\mu\otimes 1}^{{\rm LAT}}=0$ for naive and HYP-smeared quarks
(see Table~\ref{tab:CLAT}) because the taste-singlet vector bilinear
is the conserved current. This is not the case for the asqtad action
(due to the distance 3 Naik term), and so the constant need not (and
does not) vanish.
To give an idea of the size of the corrections, we note that on the
coarse MILC lattices, the momenta within
the window where NPR can be used range roughly from
$\mu = 1.0/a \approx 1.7\;$GeV to
$\mu \approx 3\;$GeV, so that
$4\pi/\alpha(\mu)$ ranges from $38-52$.
Thus one needs perturbative coefficients $C$ and $\delta$
to have magnitudes $\lesssim 10$ to have reasonable
convergence. We see from the tables that this
is the case except for the scalar (and pseudoscalar) bilinears
with the naive staggered action and operators [case (a)].
This is one of the reasons why we do not present numerical results
for this case in the main text.
The constants are smallest for HYP-smeared operators, suggesting
the PT should be better behaved in these cases.
We also note that, while mean-field improvement reduces the
magnitude of the corrections for the HYP-smeared action and
operators, this is not uniformly the case for the asqtad action,
where for scalars corrections are increased.

We close this section by describing two ways of rewriting the
perturbative results that might have some practical utility.
The first involves ratios of the differences of the initial
ratios from unity:
\begin{eqnarray}
\frac{
\left(\frac{Z_{S\otimes F1}(\mu,a)}{Z_{S\otimes I}(\mu,a)} 
\tilde u_0^{-|S|+|S-F1|}-1\right)}{
\left(\frac{Z_{S\otimes F2}(\mu,a)}{Z_{S\otimes I}(\mu,a)} 
\tilde u_0^{-|S|+|S-F2|}-1\right)}
&=&
\frac{\delta^{S\otimes F1}_{S\otimes I}}
     {\delta^{S\otimes F2}_{S\otimes I}} + {\cal O}(\alpha^2)
\,.
\end{eqnarray}
The utility of this double ratio is that
the coupling constant cancels in the 1-loop contribution,
so one obtains a simple numerical prediction.
In practice, however, there are two difficulties:
the $(a\mu)^2$ discretization errors need not cancel,
and the relative size of
the ${\cal O}(\alpha^2)$ contributions are typically
different for tastes $F1$ and $F2$.
Because of these difficulties, we have found it more useful
to simply compare the initial single ratios to PT.

An alternative way of presenting PT results for ratios
is to define $\alpha_{\rm eff}$ as follows:
\begin{equation}
\frac{Z_{S\otimes F}(\mu)}{Z_{S\otimes I}(\mu)} 
\tilde u_0^{-|S|+|S-F|}-1
= \frac{\alpha_{\rm eff} }{4\pi} \delta^{S\otimes F}_{S\otimes I}
+ {\cal O}(\alpha^2)
\,.
\label{eq:alphaeff}
\end{equation}
If the one-loop results gave a perfect representation of
the data, $\alpha_{\rm eff}$ would be the almost the same for all ratios
and independent of $\mu$. 
There would be some variation since $\alpha_{\rm eff}$ is the
coupling evaluated at a scale which we know to be of ${\cal O}(1/a)$ 
but whose precise value varies between ratios.
It would then be interesting to take the ratio of the values of
$\alpha_{\rm eff}$
at our two different lattice spacings, since this should lie in the range
$\alpha(1/a_{\rm coarse})/\alpha(1/a_{\rm fine})=1.22$ to
$\alpha(2/a_{\rm coarse})/\alpha(2/a_{\rm fine})=1.15$.
In forming this ratio one should work at fixed
$a\mu$ (rather than at fixed $\mu$), so as to better
cancel lattice artefacts.

Again, in practice we have found that the combination of
non-canceling discretization errors and taste-dependent higher-order
corrections makes this method difficult to use quantitatively.
Thus in the main text we make a more qualitative comparison between
the results on the two lattice spacings.

\section{Continuum perturbative results}
\label{app:running}

In this appendix we collect the results from the literature
that allow us to predict the matching factors using perturbation theory
using the master formula Eq.~(\ref{eq:Zmaster}).

There are many ways of writing the matching factor, with or without
intermediate schemes, and with the running over the large range of scales
taking place in different schemes. We have chosen the specific form
(\ref{eq:Zmaster}) for the following reasons.
First, by doing the running from $\mu$ to $\mu_0\approx1/a$ first,
we can, if we wish, move the running to the other side of the equation,
and so convert the lattice results into a scale-independent form.
Second, we need to use the intermediate $\overline{\rm MS}$ scheme
because the matching to the lattice is only available in this scheme
(as discussed in the previous appendix).

When evaluating the master expression, we have used the highest order
available in the literature for each part. 
For the beta-function in the $\overline{\rm MS}$ scheme, 
in the convention where
\begin{equation}
\beta(\lambda) = \frac{d\lambda}{d\ln(\mu^2)}
= - \beta^{(0)} \lambda^2-\beta^{(1)} \lambda^3 - \dots,
\end{equation}
we have (setting here, and in the following, $N_c=N_f=3$)
\begin{equation}
\beta^{(0)}=9\,,\
\beta^{(1)}=64\,,\
\beta^{(2)}=643.83\,,\
\beta^{(3)}=12090.4\,.
\end{equation}
For the anomalous dimensions in the RI$'$ scheme,
whose perturbative expansion we define as
\begin{eqnarray}
\gamma_{\CO}(\lambda) &=& - \frac{d\ln(Z_\CO)}{d\ln(\mu^2)}
\\
&=& \gamma_\CO^{(0)}\lambda + \gamma_\CO^{(1)}\lambda^2 + \dots,
\end{eqnarray}
the coefficients are known to four loops for the scalar~\cite{Chetyrkin99}
\begin{equation}
\begin{split}\gamma_S^{(0)}&=-4\,,\
\gamma_S^{(1)}=-108.67\,,\
\\
\gamma_S^{(2)}&=-3576.95\,,\
\gamma_S^{(3)}=-147207\,,
\end{split}
\end{equation}
and three loops for the tensor~\cite{Gracey03}
\begin{equation}
\gamma_T^{(0)}=1.33\,,\
\gamma_T^{(1)}=34.44\,,\
\gamma_T^{(2)}=976.64\,.
\end{equation}
The vector current also has a non-vanishing anomalous
dimension in the $RI'$ scheme, which we determine below.

The conversion factors from RI$'$ to $\overline{\rm MS}$ can be
obtained for the scalar bilinear from Ref.~\cite{Chetyrkin99} 
and for the tensor from Ref.~\cite{Gracey03}.
The results are
\begin{eqnarray}
Z^{RI',\overline{\rm MS}}_S(\mu_0)&\approx&
1-5.33 \lambda -121.37 \lambda^2 -3564.54 \lambda^3 \,,
\\
Z^{RI',\overline{\rm MS}}_T(\mu_0)&\approx&
1 + 35.07 \lambda^2 + 1207.96 \lambda^3 
\,,
\end{eqnarray}
with the coupling constant evaluated at scale $\mu_0$.

To obtain the result for the vector bilinear, we first note
that $Z_V^{{\rm RI},\overline{\rm MS}}=1$, as shown in Ref.~\cite{NPR}.
Here RI refers to the original regularization independent scheme of
Ref.~\cite{NPR},
in which the condition used to determine $Z_q$ from the
quark propagator differs from that in the RI$'$ scheme.
The condition determining $Z_q/Z_V$, Eq.~(\ref{eq:NPRvertex}),
is, however, the same in both schemes, 
from which we learn that~\cite{Huber10}
\begin{equation}
Z_V^{{\rm RI}',{\rm RI}} = Z_q^{{\rm RI}',{\rm RI}}
\,.
\end{equation}
Combining these results we find the desired matching factor
\begin{eqnarray}
Z_V^{{\rm RI}',\overline{\rm MS}} 
&=&
Z_V^{{\rm RI}',{\rm RI}} Z_V^{{\rm RI},\overline{\rm MS}}
\\
&=& 
Z_q^{{\rm RI}',{\rm RI}}
\\
&=& 1 + c_2 \lambda^2 + c_3 \lambda^3 + \dots
\\
&\approx& 1 + 9.17 \lambda^2 + 342.01 \lambda^3 
\,,
\end{eqnarray}
where the numerical values are from Ref.~\cite{Huber10}.
This in turn can be used to obtain the anomalous dimension
\begin{eqnarray}
\gamma_V^{RI'} &=& -\frac{d \ln Z_V^{{\rm RI}',\overline{\rm MS}}}
                         {d \ln(\mu^2)}
\\
&=& 2 c_2 \beta^{(0)} \lambda^3
+\left(2 c_2 \beta^{(1)} + 3 c_3 \beta^{(0)}\right) \lambda^4 
+ \dots
\\
&\approx& 165 \lambda^3 + 10407.5 \lambda^4 
\,.
\end{eqnarray}

\bibliographystyle{apsrev4-1} 
\bibliography{ref} 

\end{document}